\documentclass[final]{cvpr}

\usepackage{times}
\usepackage{epsfig}
\usepackage{graphicx}
\usepackage{amsmath}
\usepackage{amssymb}

\usepackage[dvipsnames, table]{xcolor}

\usepackage{diagbox}  % slashed cell
\usepackage{booktabs} % table
\usepackage{multirow} % table
\usepackage{subcaption} 
\usepackage{comment}
\usepackage[outline]{contour}% http://ctan.org/pkg/contour
\usepackage{wrapfig}

\usepackage[pagebackref=true,breaklinks=true,colorlinks,bookmarks=false]{hyperref}

\begin{document}

%%%%%%%%% TITLE
\title{%A Methodology and Database for the (fair/unbiased/explainable/non-semantic) Evaluation of Image Forensics Algorithms/Methods/Tools\\
%A Methodology and Database for the Non-Semantic Evaluation of Image Forensics Tools\\
Non-Semantic Evaluation of Image Forensics Tools:\\ Methodology and Database
%Non-Semantic Evaluation of Image Forensics Tools\\
%Non-Semantic Database for the Evaluation of Image Forensics Tools\\
%Non-Semantic Evaluation and Database for Image Forensics Tools\\
} %T: I like this one
%How to Evaluate Image Forensics Algorithms? A methodology and database}
% Write alternative title ideas here
% Methodology / Framework?
% Image Authentication / Image forensics / Image forgery
% Q: How to Evaluate Image Authentication Algorithms?
% Q: How Should Image Authentication Algorithms be Evaluated?
% Q: How Should we Evaluate Image Authentication Algorithms?
% T: Evaluation of image authentication algorithms: datasets and metrics

\author{Quentin Bammey, Tina Nikoukhah, Marina Gardella,\\Rafael Grompone, Miguel Colom, Jean-Michel Morel\vspace{.2em}\\
{\it Centre Borelli, \'Ecole Normale Sup\'erieure Paris-Saclay}\\
{\texttt{\small \{quentin.bammey, tina.nikoukhah, marina.gardella,rafael.grompone,}}\\ {\texttt{\small  miguel.colom-barco, rafael.grompone, jean-michel.morel\}@ens-paris-saclay.fr}}
}

\maketitle

%%%%%%%%% ABSTRACT
\begin{abstract}
With the aim of evaluating image forensics tools, we propose a methodology to create forgeries traces, leaving intact the semantics  of the image.  Thus, the only forgery cues left are the specific alterations of one or several aspects of the image formation pipeline. This methodology creates automatically forged images that are challenging to detect for forensic tools and overcomes the problem of creating convincing semantic forgeries.  Based on this methodology, we create the Trace database and conduct an evaluation of the main state-of-the-art image forensics tools.

\end{abstract}

\section{Introduction}

\begin{figure}[t]
\centering

\begin{subfigure}[t]{.49\linewidth}
\includegraphics[width=\linewidth]{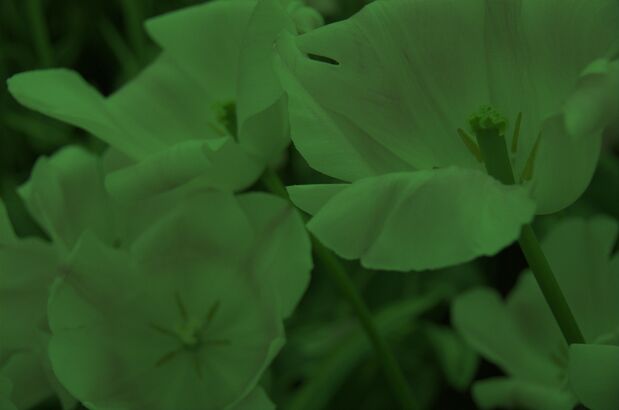}
\caption{\small Raw image}
\end{subfigure}\hfill%
\begin{subfigure}[t]{.49\linewidth}
\includegraphics[width=\linewidth]{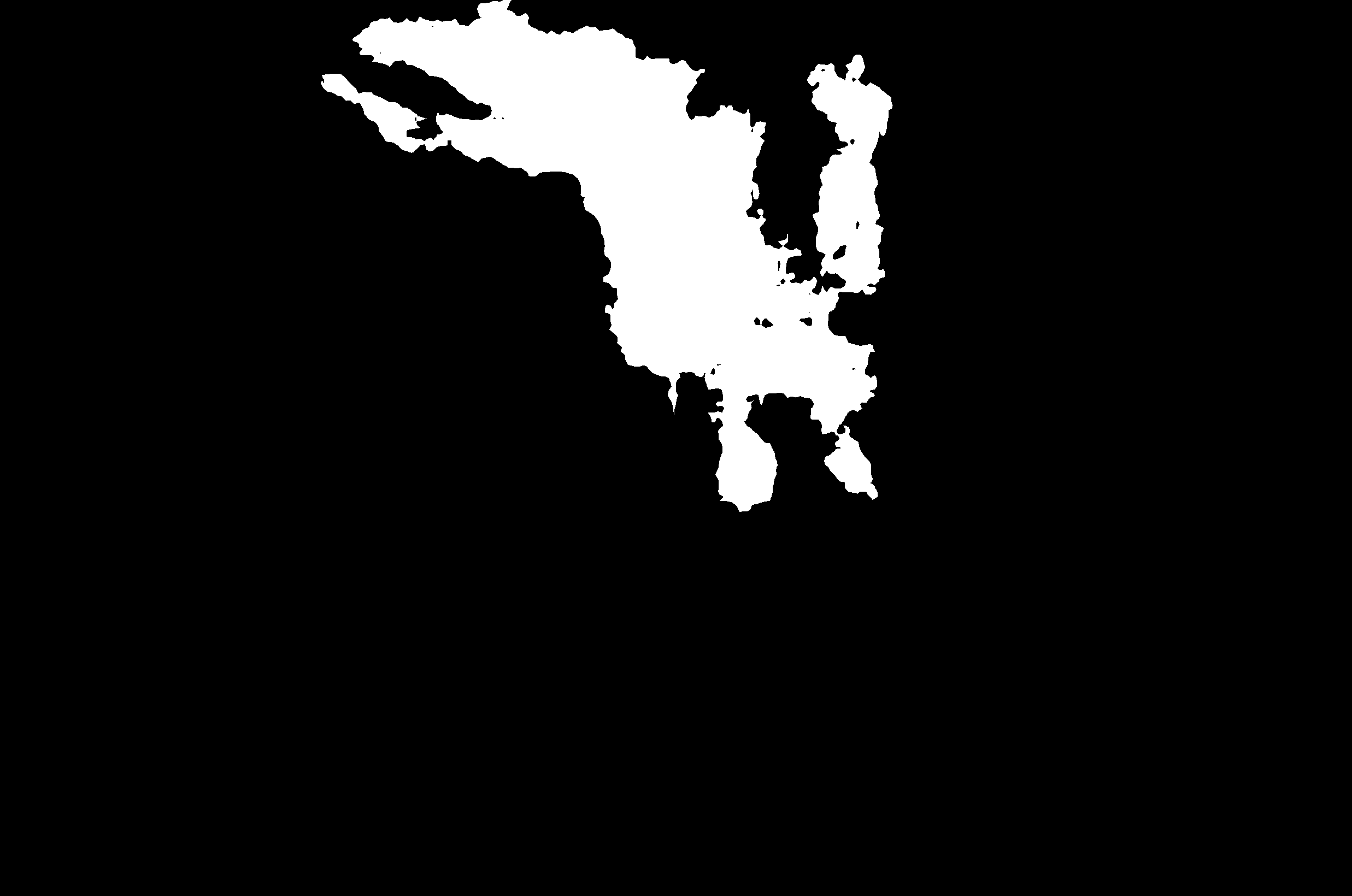}
\caption{\small Forgery mask: $M$}
\end{subfigure}

\begin{subfigure}[t]{.49\linewidth}
\includegraphics[width=\linewidth]{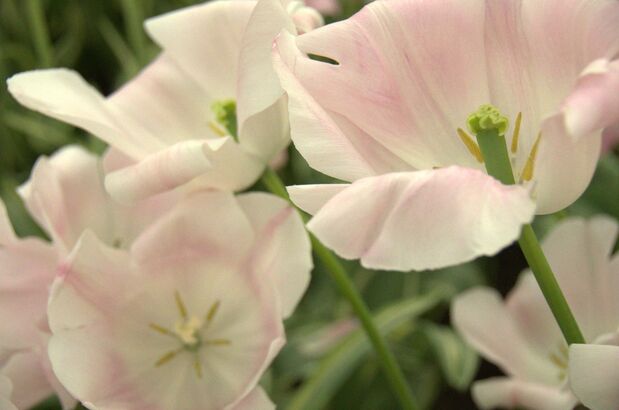}
\caption{\small Pipeline 0: $P_0$}
\end{subfigure}\hfill%
\begin{subfigure}[t]{.49\linewidth}
\includegraphics[width=\linewidth]{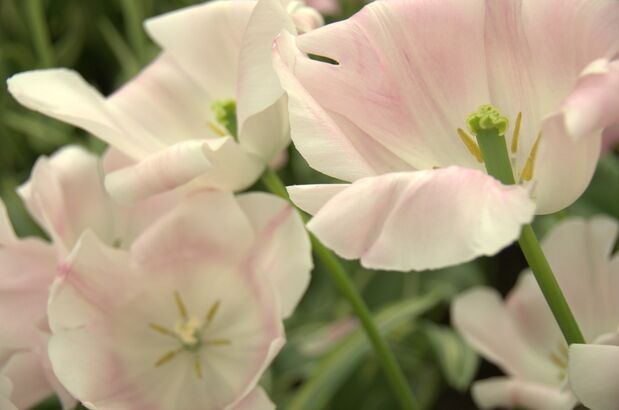}
\caption{\small Pipeline 1: $P_1$}
\end{subfigure}

\begin{subfigure}[t]{.49\linewidth}
\includegraphics[width=\linewidth]{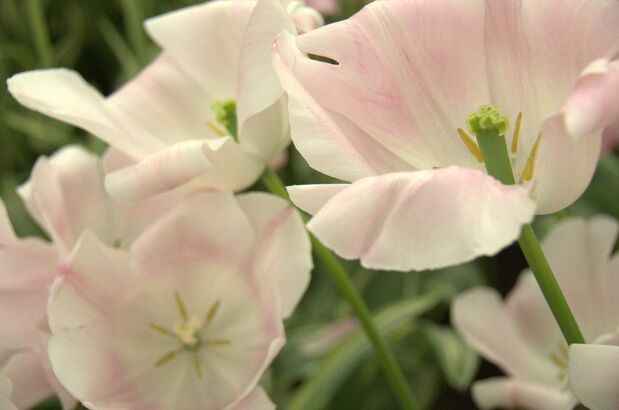}
\caption{\small Forgery: $F\!=\!\bar M\! P_0 + M\! P_1$}
\end{subfigure}\hfill%
\begin{subfigure}[t]{.49\linewidth}
\includegraphics[width=\linewidth]{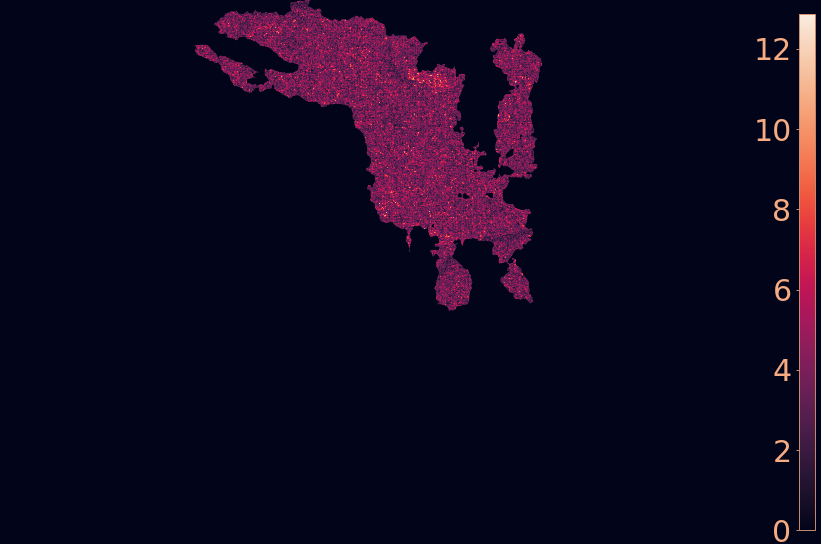}
\caption{\small Residual $|F-P_0|$}
\end{subfigure}

\begin{subfigure}[t]{.49\linewidth}
\includegraphics[width=\linewidth]{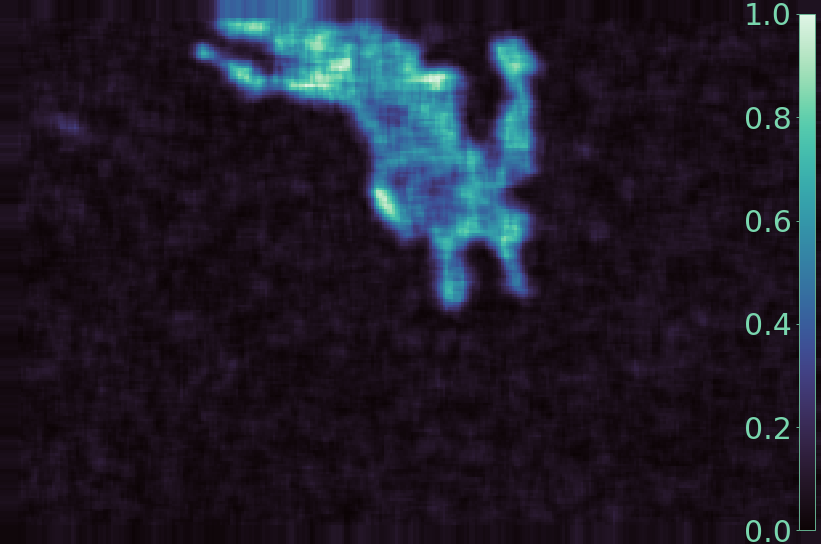}
\caption{\small Noiseprint~\cite{noiseprint} result}
\end{subfigure}\hfill%
\begin{subfigure}[t]{.49\linewidth}
\includegraphics[width=\linewidth]{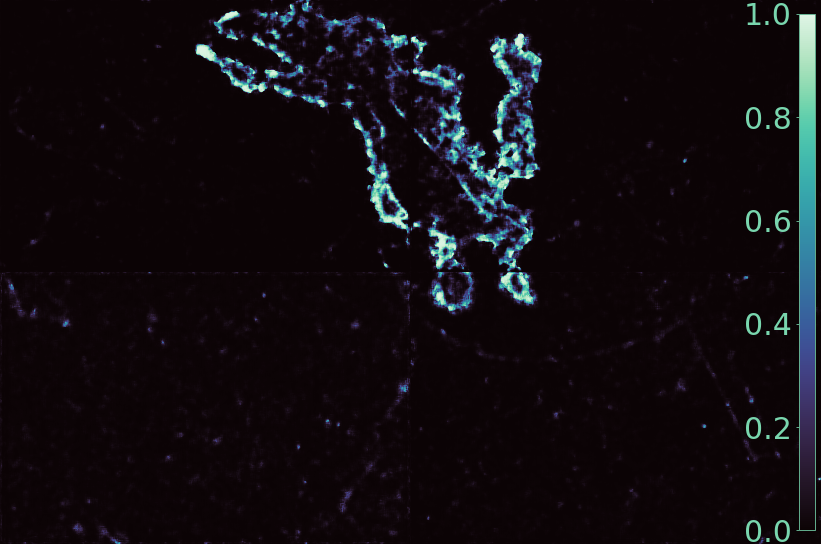}
\caption{\small ManTraNet~\cite{mantranet} result}
\end{subfigure}
\caption{%Proposed methodology for generating evaluation images.
Different image formation pipelines are applied to the same RAW image to obtain two images. The forged image is obtained combining those images using a mask.  The only difference between the authentic and forged regions are the camera pipeline traces.
%The RMSE in the forged area is 4.46.
The last row shows the result of two forensic tools applied on the forgery.}%\textcolor{red}{Redo with image r072421a1t, hybrid a.png}}
\label{teaser}
\vspace{-10pt}
\end{figure}

\iffalse % old figure
\begin{figure}[t]
\centering
\def\s{0.22\textwidth}
\begin{tabular}{@{}c@{\hskip 1em}c@{}}
\includegraphics[width=\s]{figures/teaser/down.jpeg} & 
\includegraphics[width=\s]{figures/teaser/mask.png}  \\
\small RAW image & \small $M$: Mask \\
\includegraphics[width=\s]{figures/teaser/hs0.png} & 
\includegraphics[width=\s]{figures/teaser/hs1.png}  \\
\small $P_1$: Pipeline 1 & \small $P_2$: Pipeline 2 \\
\includegraphics[width=\s]{figures/teaser/hybrid_select.png} & 
\includegraphics[width=\s]{figures/teaser/diff.jpeg}  \\
\small Forgery: $F=\bar{M}P_1+MP_2$ & \small $|F-P_1|$  \\
\includegraphics[width=\s]{figures/teaser/mahdian_h.png} & 
\includegraphics[width=\s]{figures/teaser/adq2_h.png}  \\
\small Mahdian~\cite{mahdian2009} result & \small I-CDA~\cite{ADQ2} result
\end{tabular}
\caption{%Proposed methodology for generating evaluation images.
Two image formation pipelines are applied to the RAW image to obtain images $P_1$ and $P_2$.  The forged image $F$ is obtained combining image $P_2$ inside the mask $M$ and the image $P_1$ otherwise.  The only difference between the authentic and forged regions are the image traces. The RMSE in the forged area is 4.96.  The last row shows the results of two forensic tools applied on $F$.}
\label{teaser}
\vspace{-10pt}
\end{figure}
\fi

Digital images play an extensive role in our lives and forgeries are present everywhere \cite{farid_book}. Many image processing tools are available to create visually realistic image alterations.  Yet these modifications leave behind cues: each operation has an impact on the image in the form of a particular trace. A first class of forgery detection tools aims at detecting these traces in a suspicious image by finding  local inconsistencies. Other forgery detection tools are generic, and directly trained on databases containing forged images.
A semantic analysis of an image can provide hints, but the rigorous proof of a forgery should not be based purely on semantic arguments.
%, because to refute the semantics of an image often amounts to refuting its purpose. 
%\jm{I like the next paragraph ending at "not on semantic arguments" but I think that putting so early in the introduction amounts to a distraction. It should rather be put in the final discussion. Indeed, the introduction should be focused on the state of the art and ending by our claims. The conclusion is indeed debatable: if we have an image of Joe Biden shaking hands with, say, Mao Zedong, a semantic argument will seem  sufficient to most people. On the other hand, even in that obvious case we want to argue that it remains important to make a non semantic analysis. But, again, this discussion makes more sense at the end of the paper.}\qb{Resolved during meeting: keep it like this as it removing it from the introduction would require a complete change of the article's structure}
The situation is similar to the dilemma arising from the observations of Galileo, which contradicted the accepted knowledge of his time.  In the words of Bertolt Brecht~\cite{brecht}:
\begin{quote}\small
\textsc{Galileo}: How about your highness now taking a look at his impossible and unnecessary stars through this telescope?\\
\textsc{Mathematician}: One might be tempted to answer that, if your tube shows something which cannot be there, it cannot be an entirely reliable tube, wouldn't you say?
%\textsc{Galileo}: How would it be if your Highness were now to observe these impossible as well as unnecessary stars through this telescope?\\
%\textsc{The Mathematician}: One might be tempted to reply that your telescope, showing something which cannot exist, may not be a very reliable telescope, eh?
\end{quote}
\clearpage

The telescope could have been unreliable, indeed, and a scientific inquiry on the instrument could have been justified.  However, concluding like the Mathematician that the telescope was unreliable \emph{just} based on the contents of the observations is not prudent.  Similarly, the proof of a forgery must be based on image traces, not on semantic arguments, because the \textbf{semantics of an image are usually the purpose and not the means of a forgery}.

Image forensics algorithms  are mainly  evaluated by their performance in benchmark challenges. This practice has several limitations: in many cases, the same database is split into training and evaluation data. As a consequence, algorithms are trained and evaluated on images that have gone through   similar image processing pipelines, forgery algorithms and anti-forensic tools. Hence, there is no guarantee that such learning-based methods will work in the wild, where those parameters vary much more.  

Regardless of the variety of the training set, the question arises of  whether the forgeries are being detected by trained detectors for semantic reasons, or because of local inconsistencies in the image.

With these considerations in mind, we propose a methodology and a database to evaluate image forensic tools on images where authentic and forged regions only differ in the traces left behind by the image processing pipeline. Using this methodology, we create the Trace database by adding various forgery traces to raw images from the Raise~\cite{raise} dataset. See Fig.~\ref{teaser}.  This procedure avoids the difficulties of producing convincing and unbiased semantic forgeries, which often requires manual work.

This paper is organised as follows. The next section discusses related works.  Then, Section~\ref{sec:pipeline} gives a brief description of the image formation pipeline and the main traces of each step.  Section~\ref{sec:database} gives the details of the generation of the proposed non-semantic database, which is used to evaluate state-of-the-art image forensic tools in Section~\ref{sec:experiments}.   Section~\ref{sec:dicussion} is a general discussion.

\section{Related Works}\label{sec:related-work}

There is a large literature on image forensics, starting from the seminal work of Farid~\cite{farid_book}.  Some of the methods  focus on the detection of a specific tampering technique such as copy-move or splicing. But the most classic forgery-detection methods aim at detecting local perturbations of the traces left in the image by the processing chain. Such local disruptions hint at a local forgery.
To do so, these methods strive to suppress image content and highlight intrinsic artefacts left by  demosaicking,  JPEG encoding, etc.~\cite{Popescu2004StatisticalTF}. Hence these forgery detection methods can be grouped by their specifically-targeted artefacts, which we now briefly review.% in consonance with the specifically targeted artefact. 

%Despite the fact that most datasets classify forged images according to the tampering technique being used, most of the classical forgery-detection methods aim to detect underlying inconsistencies left by the forgers rather than a specific forgery type. To do so, these methods strive to suppress image content and highlight intrinsic artefacts~\cite{Popescu2004StatisticalTF}. In this sense, forgery-detection methods can be grouped in consonance with the specific artefact they target. 

Noise-based methods try to reveal local inconsistencies in noise models (see Section~\ref{sec:pipeline}) that could result from tampering. The method proposed in~\cite{mahdian2009}, consists in performing local wavelet based noise level estimation using a median absolute deviation estimator. In~\cite{lyupan2013}, noise estimation is done based on the kurtosis concentration phenomenon. Both methods deliver a heat-map in which regions showing a different noise level are pointed out as suspicious. A still more sophisticated approach in~\cite{splicebuster2015} uses the co-occurences of noise residuals as local features  revealing tampered image regions.

%\begin{description}

%\item [Mahdian] This method, proposed by Mahdian and Saic~\cite{mahdian2009}, consists in performing local wavelet based noise level estimation using a MAD (\textit{median absolute deviation}) estimator. Regions of an image showing a different noise level are pointed out as suspicious.

%\item [Lyu] This method, proposed by Lyu et al.~\cite{lyupan2013} estimates local noise variances based on the observation that kurtosis values of natural images across different band-passed filter channels concentrate around a constant value. Regions of an image showing a different noise level are pointed out as suspicious.

%\item [Splicebuster] This algorithm, proposed by Cozzolino et al.~\cite{splicebuster2015} performs local feature extraction based on noise residuals and their co-occurrence histograms. The distribution of these histograms is assumed to follow a mixture model of two classes whose parameters are estimated using the expectation-maximisation algorithm. The final output shows, for each analysed block, the ratio between its distance to both classes.

%\end{description}

   Detecting the specific image demosaicking algorithm (see Section~\ref{sec:pipeline}) has not been attempted since the 2005 pioneer paper by Popescu and Farid~\cite{popescu_cfa}, conceived at a time where those algorithms were simpler and easier to distinguish. However, detecting the mosaic pattern has received more extensive coverage. Choi et al.~\cite{choi} used the fact that sampled pixels were more likely to take extremal values, while Shin et al.~\cite{shin2017color} noticed that they had a higher variance. More recently, Bammey et al.~\cite{bammey} combined the translation invariance of convolutional neural networks with the periodicity of the mosaic pattern to train a self-supervised network into implicitly detecting demosaicing artefacts.

The traces left by JPEG compression are blocking effects and quantization of the DCT coefficients of each image block. One can divide the JPEG forensic tools into two categories. BAG~\cite{BLK} and CAGI~\cite{CAGI} analyse blocking artefacts, while other methods analyse the DCT coefficients. More precisely, CDA~\cite{ADQ1} and I-CDA~\cite{ADQ2} are based on the AC coefficient distributions, while FDF-A~\cite{ADQ3} is based on the first digit distribution of AC coefficients. Zero~\cite{ZERO} counts the number of  null DCT coefficients in all blocks  and deduces the grid origin. 

More recently, generic tools were proposed based on neural networks.  Noiseprint~\cite{noiseprint}  uses a Siamese network trained on authentic images to extract a local fingerprint. Using this network on different patches of an image enables it to decide whether the two patches come from the same camera or not, which can lead to detect splicing attacks.
 ManTraNet~\cite{mantranet} is an end-to-end network  with two parts: the first part is trained to detect image-level manipulations, while the second part is trained on synthetic forgery datasets to detect and localise forgeries in the image.
Finally, the Self-consistency~\cite{selfconsistency} method also uses a Siamese network with the goal of detecting whether two patches have been processed with the same pipeline. 
They make use of N-Cuts segmentation~\cite{ncuts} to automatically cluster and detect relevant traces of forgeries.

%\begin{description}
%\item [Noiseprint~\cite{noiseprint}] This method trains a Siamese network on authentic images to extract a local signature of the image that identifies the camera. Using this network on different patches of an images enables it to decide whether the two patches come from the same camera. \marina{I think this needs some further explanation on why it is a generic method and not a noise method. JM: it is not noise, it is a local signature that only depends on the camera, this signature is very noisy, but  this "noise" has a different structure on tampered regions than the in the rest. At the end you need to look at it. Anyway I have changed the description to avoid talking about noise. }

%\item[ManTraNet~\cite{mantranet}] This paper proposes an end-to-end network that consists of two parts. The first part is trained to detect image-level manipulations, while the second part is trained on synthetic forgery datasets to detect and localize forgeries in the image.

%\item[Self-consistency~\cite{selfconsistency}] This method uses a Siamese network with the goal of detecting whether two patches have been processed with the same pipeline. Because information about the pipeline is usually found in the EXIF metadata of authentic images, they use the EXIF metadata as a self-supervisory means of detecting whether two parts of an image are likely to have come from the same image processing pipeline. They finally make use of N-Cuts segmentation~\cite{ncuts} to automatically cluster and detect relevant traces of forgeries.

%\end{description}

There is also a considerable literature proposing datasets for the evaluation of forensic tools.  An early example is the Columbia Dataset~\cite{NgChangTR20320043}, which only contains spliced $128 \times 128$ grayscale blocks for which no masks are provided.  %Two years later, the Columbia Color Dataset~\cite{} added color images of better resolution and forgery masks.
New benchmarks were proposed in 2009 with CASIA V1.0 and V2.0~\cite{casiadataset}.  These datasets included splicing forgeries and copy-move attacks, with a total of 8000 pristine images and 6000 tampered images.  Post-processing was introduced as a counter-forensics technique.  MICC F220 and F2000 datasets~\cite{miccdataset} as well as the IMD dataset~\cite{imd2012} provide further benchmarks for copy-move forgery detection. These datasets were constructed in an automatic way. While the first two randomly select the region of the image to be copy-pasted, IMD dataset performed snippets extraction. %Both approaches include the possibility of adding several artefacts to the forged region.
Other datasets adressing copy-move forgeries with post-processing counter attacks are also available~\cite{comofod2013, 7532339}.
Image forgery-detection challenges are another source of benchmark datasets. The National Institute of Standards and Technology (NIST) organizes, since 2017, an annual challenge for which different datasets are released~\cite{MFCdatasets}. %Forgeries included in those datasets are of varied type and they are generated both automatically and manually. 
This includes both automatically and manually generated
forgeries.

Some datasets are built with the aim of providing forgeries  imperceptible to the naked eye. A good example is the Korus dataset~\cite{Korus2016WIFS,Korus2016TIFS} which consists of 220 pristine images and 220 handmade tampered images consisting in object removal or insertion. 
%so various regions of the background image were modified to camouflage the forgery without semantic changes. 
The recent DEFACTO dataset~\cite{DEFACTODataset} is constructed on the MSCOCO dataset~\cite{lin2015microsoft} and  includes a wide range of forgeries such as copy-move, splicing, inpainting and morphing.
%{\color{red}[The]}\qb{no: would imply that some of the forgeries are non-semantic, whereas all forgeries are semantic AND generated as described below}
Semantically meaningful forgeries are generated automatically but with several biases such as copy-pasting objects in the same axis or only performing splicing with simple objects.
Bammey et al.~\cite{bammey} %, instead of focusing on a specific forgery, instead 
proposed a dataset to compare localised detection of demosaicing forgeries by randomly merging images demosaiced in different way. %However, images created this way are semantically incorrect.
However, as a random combination of two unrelated images, the resulting forgeries present semantic incongruities, and are thus not suited to evaluate the ability of an algorithm to detect forgeries without content-awareness.
%\jm{I do not get what you mean by semantically incorrect, can you explain.}\qb{solved}
%The result is, however, semantically incorrect.

%Image forgery-detection challenges are another source of benchmark datasets. The National Institute of Standards and Technology (NIST) organizes, since 2017, an annual challenge for which different datasets are released~\cite{MFCdatasets}. Forgeries included in those datasets are of varied type and they are generated both automatically and manually. 
%Despite classical datasets which are usually classified according to the tampering technique being used, challenge datasets mix different type of forgeries and split the tampered images in training and evaluation datasets.

Most recent forgery-detection datasets start from pristine images %from the real-world 
and perform  several different sorts of forgeries on them~\cite{surveyimagetampering}. Since early datasets~\cite{casiadataset,hsu06crfcheck,NgChangTR20320043}, the number of tampering techniques has increased to include new ones such as colorization~\cite{CASTRO2020104864}, inpainting~\cite{CASTRO2020104864, DEFACTODataset} and morphing~\cite{CASTRO2020104864, DEFACTODataset, umdfaceswap}. 
Post-processing and counter-forensic techniques have been increasingly used to produce visually imperceptible forgeries; %. However,
but such post-processing may also leave detectable traces.
%Furthermore, we can observe that the use of post processing and counter forensics techniques has increased to produce visually imperceptible forgeries that better represent the real world challenges in image-forgery detection. However, the traces left by these counter forensics approaches can also be used for detection purposes. 
Efforts have also been made in order to automatically obtain large datasets. However, the resulting tampered images are either semantically incorrect~\cite{miccdataset,imd2012} or biased~\cite{DEFACTODataset}. Both scenarios pose problems for training neural networks, which risk 
overfitting on the forgeries' methods and semantic content.
%either overfitting or semantically detecting forgeries.
%\jm{why "semantically detecting" would be a defect? If it is, this must be that such detection is fitted to the database and not generalizable. So this deserves explanation.}\qb{better like this?}
%: in the first case, these algorithms could be learning to detect forgeries in a semantical way, while in the second case they risk overfitting.

\section{Image formation pipeline}\label{sec:pipeline}

\begin{figure*}[t]
    \centering
    \includegraphics[width=\textwidth]{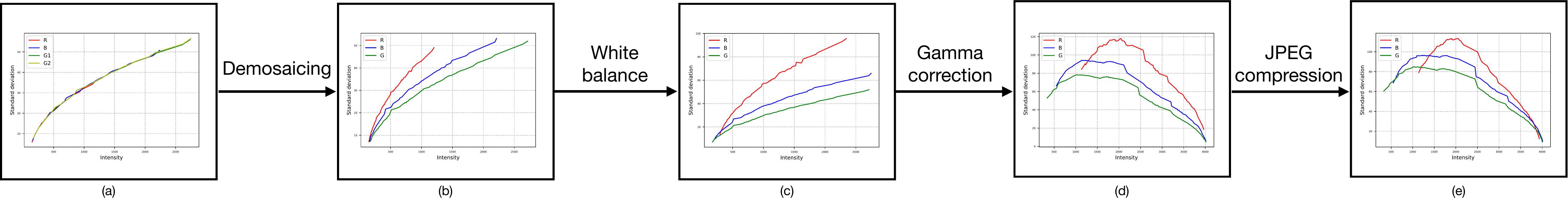}
    \caption{Image processing pipeline, and evolution of the noise curves throughout the different steps. Complete pipeline for ISO 1250, t=$1/30s$, Canon EOS 30D~\cite{mcolom_phd, colom2014nonparametric, 7113861}.}%: raw image, demosaicing, white balance, gamma correction, and JPEG compression. In the first step (raw image), all four channels share the same noise curve. After demosaicing, each channel has a different noise curve since the Adams-Hamilton algorithm treats each channel differently. Gamma correction saturates the noise curve, which starts to decrease after a certain intensity. Finally, noise is reduced after JPEG compression due to the quantization of DCT coefficients~\cite{mcolom_phd, colom2014nonparametric, 7113861}.}
    \label{fig:complete_pipeline_canon_1250}
    \vspace{-10pt}
\end{figure*}

%In this section we will make a brief description of the principal steps of the camera processing chain. %Other important steps, such as denoising, will be omitted since they are out of the scope of this paper.
Figure~\ref{fig:complete_pipeline_canon_1250} summarises the image processing pipeline~\cite{delbracio2021mobile} and shows how the noise curves change at its different steps.
%shows the evolution of the noise curves along the complete camera processing chain, for ISO 1250, t=$1/30s$ exposure time, Canon EOS 30D camera ~\cite{mcolom_phd, colom2014nonparametric, 7113861}.

\vspace{-10pt}
\paragraph{Raw image acquisition} 
%The first step to acquire a digital image is to transform light radiation into an electrical signal. To do so, cameras have sensors that count the number of incident photons along the exposure time and transform it into voltage values. The final step consists in the conversion of the analog voltage measures into digital quantized values.  

%Different technologies can be used in sensors, being charge-coupled devices (CCD) and complementary metal–oxide–semiconductors (CMOS) the two most currently used.
%Although digital devices having three sensors exist, most digital cameras are equipped with a single sensor array that is unable to separate color information. In order to obtain a color image, a color filter array (CFA) is placed in front of the sensor to split incident light components according to their wavelength.% This way, each cell measures the intensity of a pixel with respect to a given color. 

The value at each pixel %generated by the process above
can be modelled as a Poisson random variable whose expectation is the real pixel value~\cite{Foi2008}. Therefore, noise variance at this step follows a simple linear relation $\sigma^2 = A + Bu$ where $u$ is the intensity of the ideal noiseless image and $A$ and $B$ are constants (see Fig.~\ref{fig:complete_pipeline_canon_1250}(a)). Furthermore, given the nature of the noise sources at this step, noise can be accurately modelled as uncorrelated, meaning that noise at one pixel is not related with the noise at any other pixel.

%Figure~\ref{fig:complete_pipeline_canon_1250} shows the evolution of the noise curves along the complete camera processing chain, for ISO 1250, t=$1/30s$ exposure time, Canon EOS 30D camera. It shows the raw image, demosaicing, white balance, gamma correction, and JPEG compression. In the first step (raw image), all four channels share the same noise curve. After demosaicing, each channel has a different noise curve since the Adams-Hamilton algorithm treats each channel differently. Gamma correction saturates the noise curve, which starts to decrease after a certain intensity. Finally, noise is reduced after JPEG compression due to the quantization of DCT coefficients.~\cite{mcolom_phd, colom2014nonparametric, 7113861}.

\vspace{-10pt}
\paragraph{Demosaicing}
%\marina{To be modified by Quentin}

Most digital cameras are equipped with a single sensor array that is unable to separate colour information. In order to obtain a colour image, a colour filter array (CFA) is placed in front of the sensor to split incident light components according to their wavelength. The raw image obtained from the sensor has one colour component per pixel: red, green, or blue. The demosaicing process consists in the reconstruction of a full colour image from the incomplete colour samples by interpolating the two missing colour values per pixel.  Figure~\ref{fig:complete_pipeline_canon_1250}(b) shows that after demosaicing, each channel has a different noise curve since channels are processed differently by the demosaicing algorithm. Furthermore, noise is no longer uncorrelated due to the fact that demosaicing algorithms use information of nearby pixels to interpolate missing values. 

\vspace{-10pt}
\paragraph{White Balance}

In order to obtain a faithful representation of the colours as perceived by the observer, colour intensities are adjusted in such a way that achromatic objects from the real scene are rendered as so~\cite{losson:hal-00705825}. This process %of color calibration is commonly 
is known as white balance and consists in scaling each channel's value by multiplying it by a constant.  The effects of white balance in noise are shown in Figure~\ref{fig:complete_pipeline_canon_1250}(c). The multipliers are always bigger than one, so the noise is increased after this step. Since channels have different multipliers, noise increments are different for each one.

\vspace{-10pt}
\paragraph{Gamma Correction}

Given that the relationship between stimulus and human perception is not linear but rather logarithmic~\cite{fechner1860elemente}, cameras process the intensity of each channel by applying a power law function of the form $G_{k, \gamma}(u) = k u^{\gamma}$.  %Due to the power law function,
As a consequence, noise is significantly increased %after gamma correction,
as it is shown in Figure~\ref{fig:complete_pipeline_canon_1250}(d). However, the main change is that the noise curves are no longer monotonically increasing after the gamma correction.

\vspace{-10pt}
\paragraph{JPEG compression}

The JPEG image standard is the most popular lossy compression scheme for photographic images~\cite{wallace1992jpeg}. 
%The JPEG encoding consists first on a color space transformation and an optional sub-sampling of each channel. Then, each channel is partitioned into non-overlapping $8\times8$-pixel blocks and the orthogonal 2D discrete cosine transform (DCT) type II is applied to each of these blocks.
The image goes through a colour space transformation and each channel is partitioned into non-overlapping $8\times8$-pixel blocks. The type-II discrete cosine transform (DCT) is applied to each of these blocks.
The resulting coefficients are quantized according to a table (which depends on a quality factor Q) and the coefficients are then losslessly compressed.
%Each of the $8\times8$ blocks undergoes a quantization step performed in the spectral domain. A quantization table (related to the compression quality) provides a factor for each DCT component. At this step, some DCT coefficients are cancelled out when they have a small value relative to the quantization factor. Finally, the quantized DCT coefficients are lossless-compressed by exploiting, among other things, the presence of zero values.     
%Due to the independent encoding, $8\times8$ blocking artefacts are introduced in the decompressed image. The smaller the quality factor is, the more the image is compressed and the more evident this effect becomes.
Figure~\ref{fig:complete_pipeline_canon_1250}(e) shows that noise is reduced after JPEG compression. This is mainly because of the colour space change and the cancellation of small high-frequency DCT coefficients during the quantization step.

\section{Database}\label{sec:database}

We created a database of forgeries which leave intact the semantics of the image. The global idea of our method is to take a raw image, process it with two different pipelines, and merge the two processed images as follows: the first image is used to represent the authentic region, and the second image is used to make the forged area given by a mask, as can be seen in Fig.~\ref{teaser}.
As a base for our forgeries, we use the RAISE-1k dataset~\cite{raise}, which contains one thousand pristine raw images of varied categories, taken from three different cameras.
We note that the variety of source cameras is not important to our forgery database, as we erase the previous camera traces by downsampling the image, then resimulate the whole image processing pipeline ourselves, as explained below.
Furthermore, the code can be used with any other source images, to automatically generate large quantities of forgeries.

\vspace{-10pt}
\paragraph{Methodology for the creation of the database}
A raw image is already contaminated with noise and, furthermore, its pixels are all sampled in the same CFA pattern. In order to control the noise and CFA pattern, we start by downsampling each image by a factor 2. This enables us to choose the amount of noise to be added, and to mosaic the image in any of the four possible patterns.
Once the image has been downsampled, we process the image with two different pipelines.
The two images are then merged as explained above.% The first image serves as the authentic base image, and the second image is pasted onto the first at the location of the forgery mask, whose creation is described below.}

\vspace{-10pt}
\paragraph{Forgery masks}

\begin{figure}[t!]
    \centering
    \begin{tabular}{@{}c@{\,}c@{}}
    \includegraphics[width=.49\linewidth]{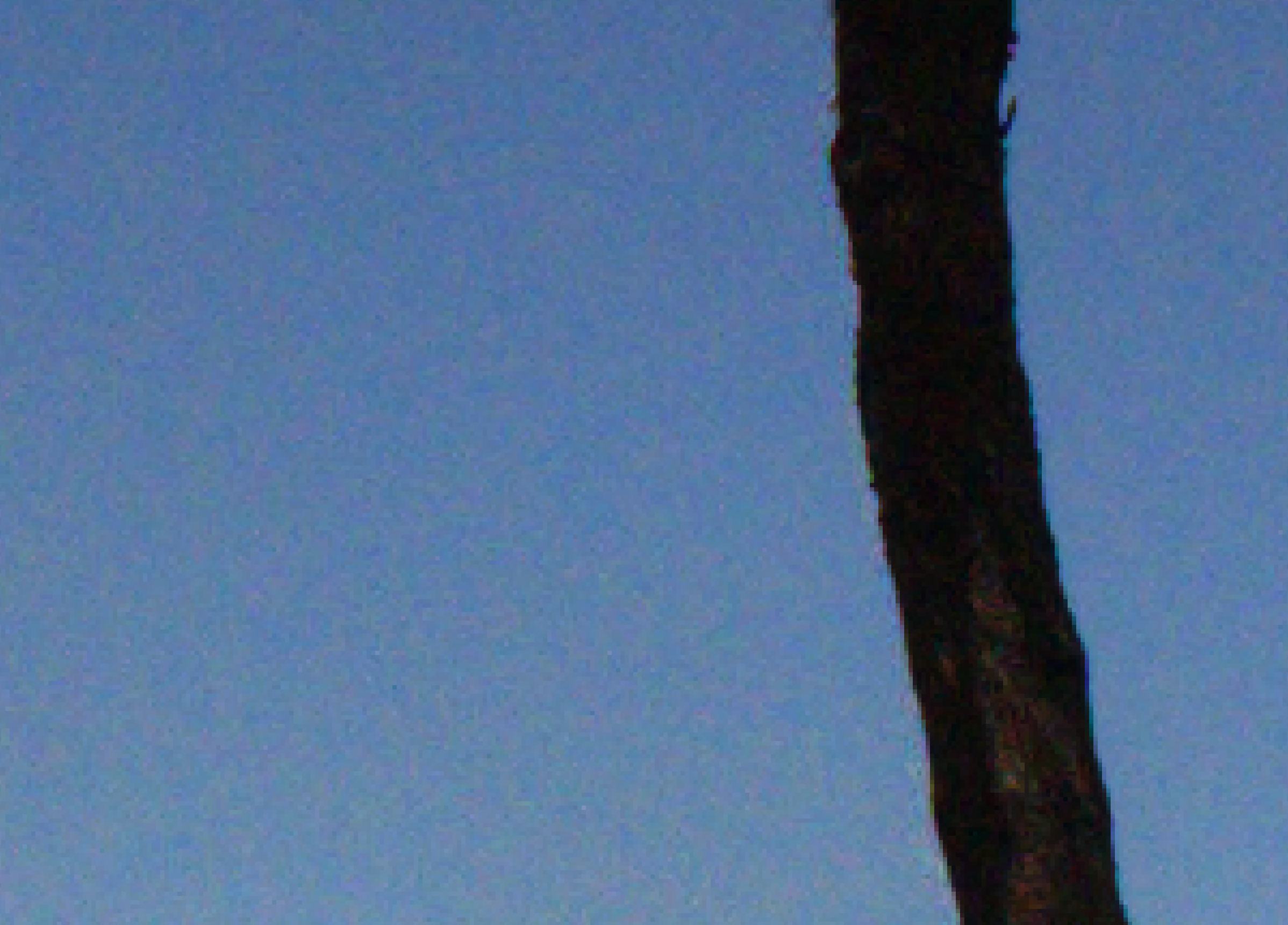}&%
    \includegraphics[width=.49\linewidth]{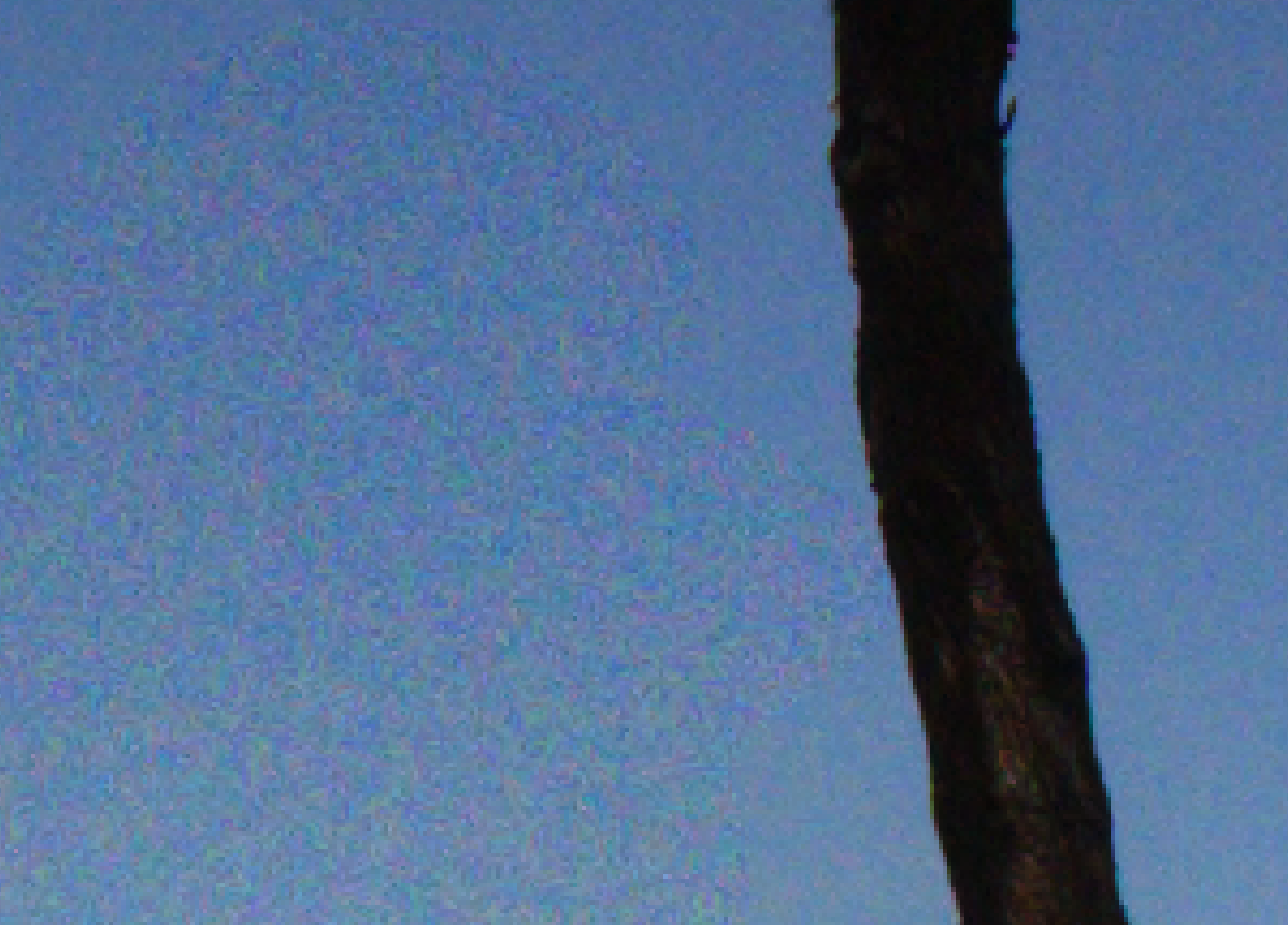}\\
    \includegraphics[width=.49\linewidth]{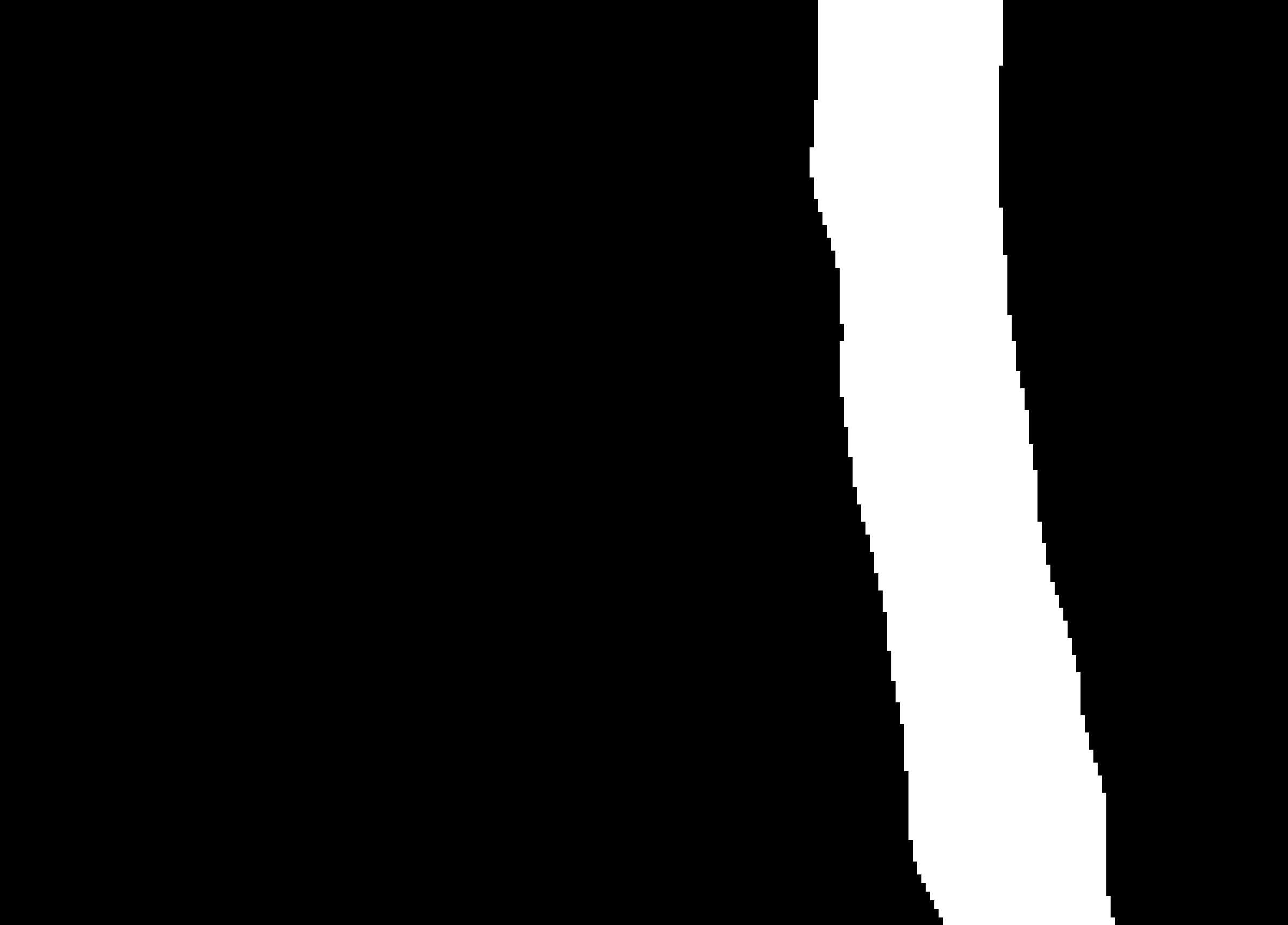}&%
    \includegraphics[width=.49\linewidth]{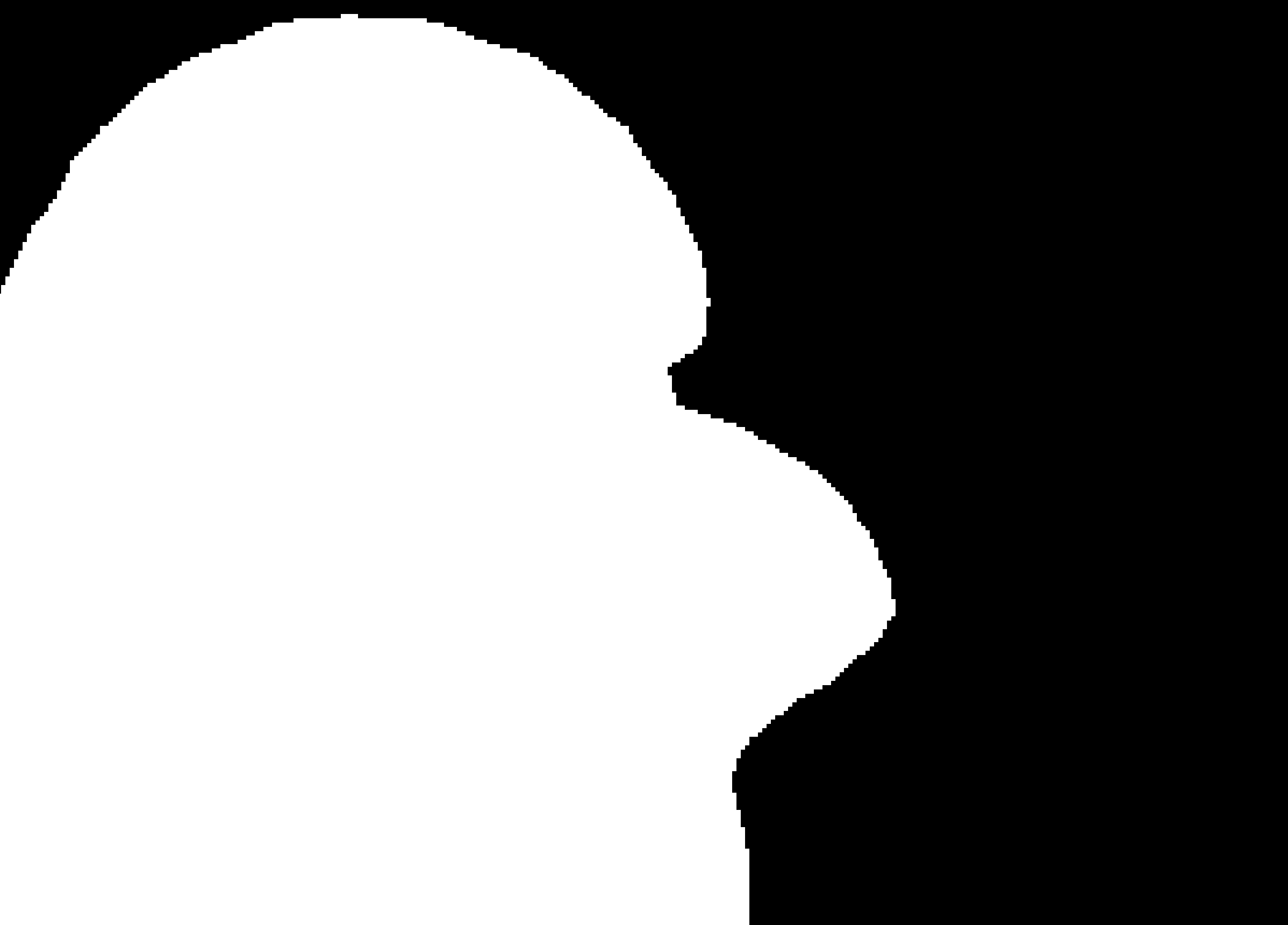}
    \end{tabular}
    \caption{Zoom of the same image, with forgeries made using the two masks. On the left, the endomask coincides with the image's structure, here a tree. The forgery is less conspicuous than on the right where the exomask is in the sky, where the borders do not coincide with the images' content. 
     }%\tina{r1b09da9ct}
    \label{fig:newmask}
    \vspace*{-2.5mm}
\end{figure}

For each image we construct two different masks. 
Since inconsistencies in the image processing pipeline are usually most visible at the border of the forgery, the first mask is constructed in order to coincide to an object in the image. To do this, we segment the original images with EncNet~\cite{encnet}. For each image, we take a pixel at random, and select the connected component it belongs to. We accept the mask if its size is less than half the image's, otherwise we pick another pixel until we find a suitable mask. This ensures that each image has only one forgery, whose size is at most half the image's. Using masks that correspond to a component of the image segmentation ensures almost invisible forgeries, since the borders will match the image's structure, as shown in Fig.~\ref{fig:newmask}. These endogenous masks, or \emph{endomasks}, emerge from the image itself.

\begin{figure}
\centering
\def\s{0.33\linewidth}
\def\ss{-2pt}
\begin{tabular}{@{}c@{}c@{}c@{}}
%\toprule
Endomask & Image & Exomask\\
%\midrule
\includegraphics[width=\s]{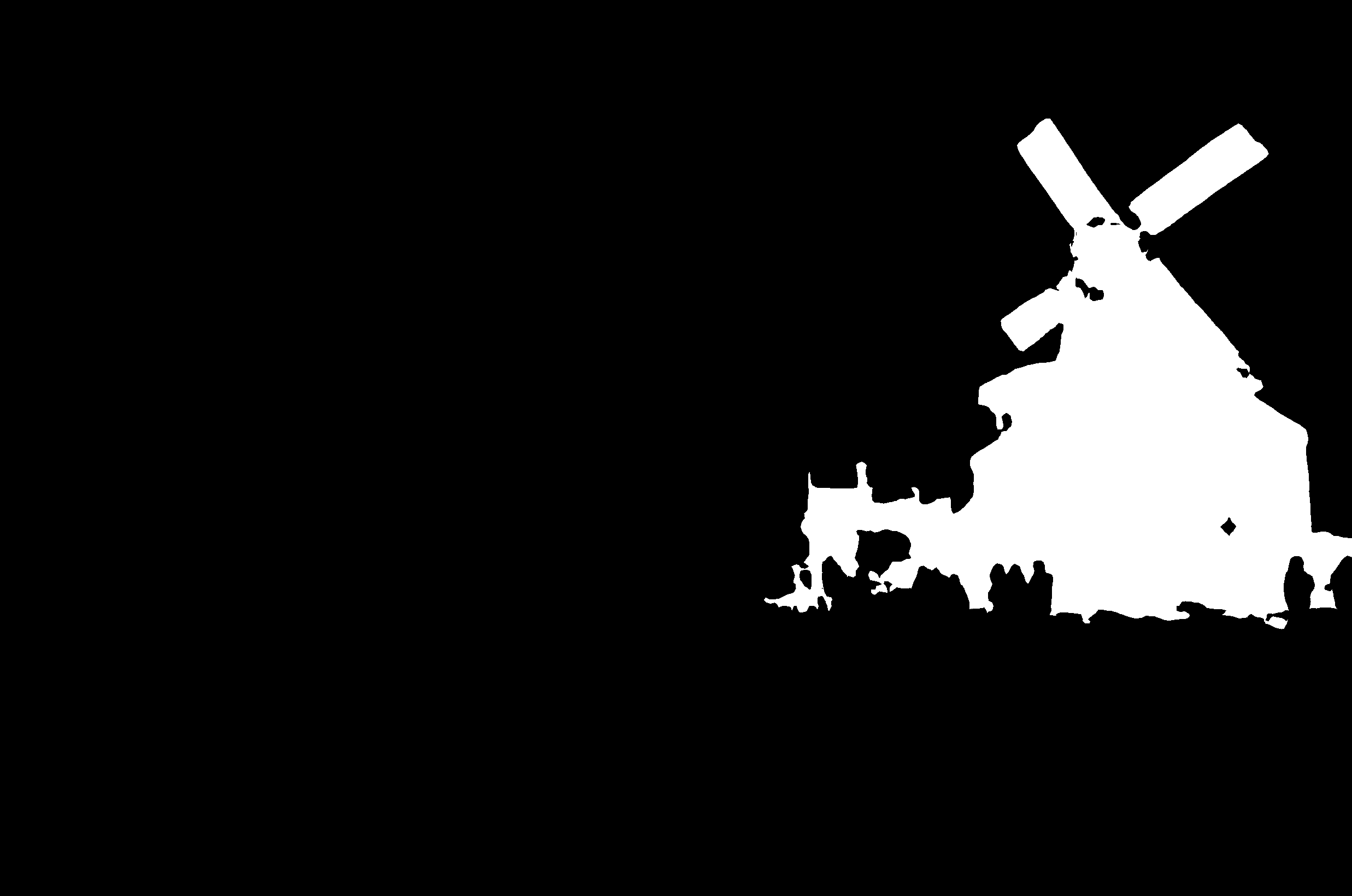}&%
\includegraphics[width=\s]{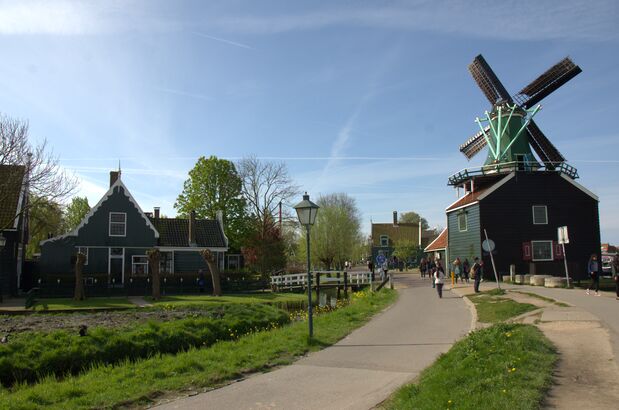}&%
\includegraphics[width=\s]{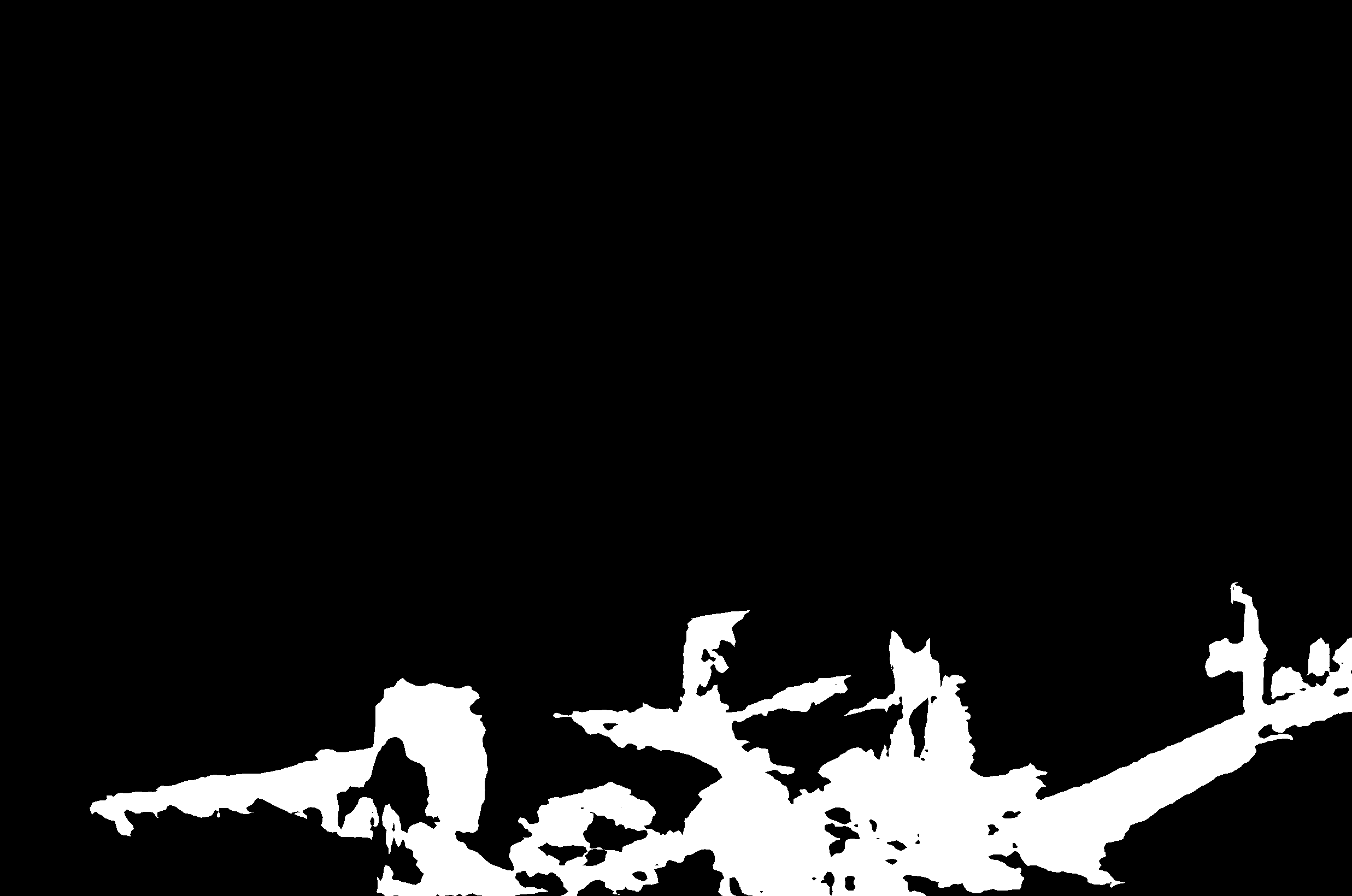}\\[\ss]
\includegraphics[width=\s]{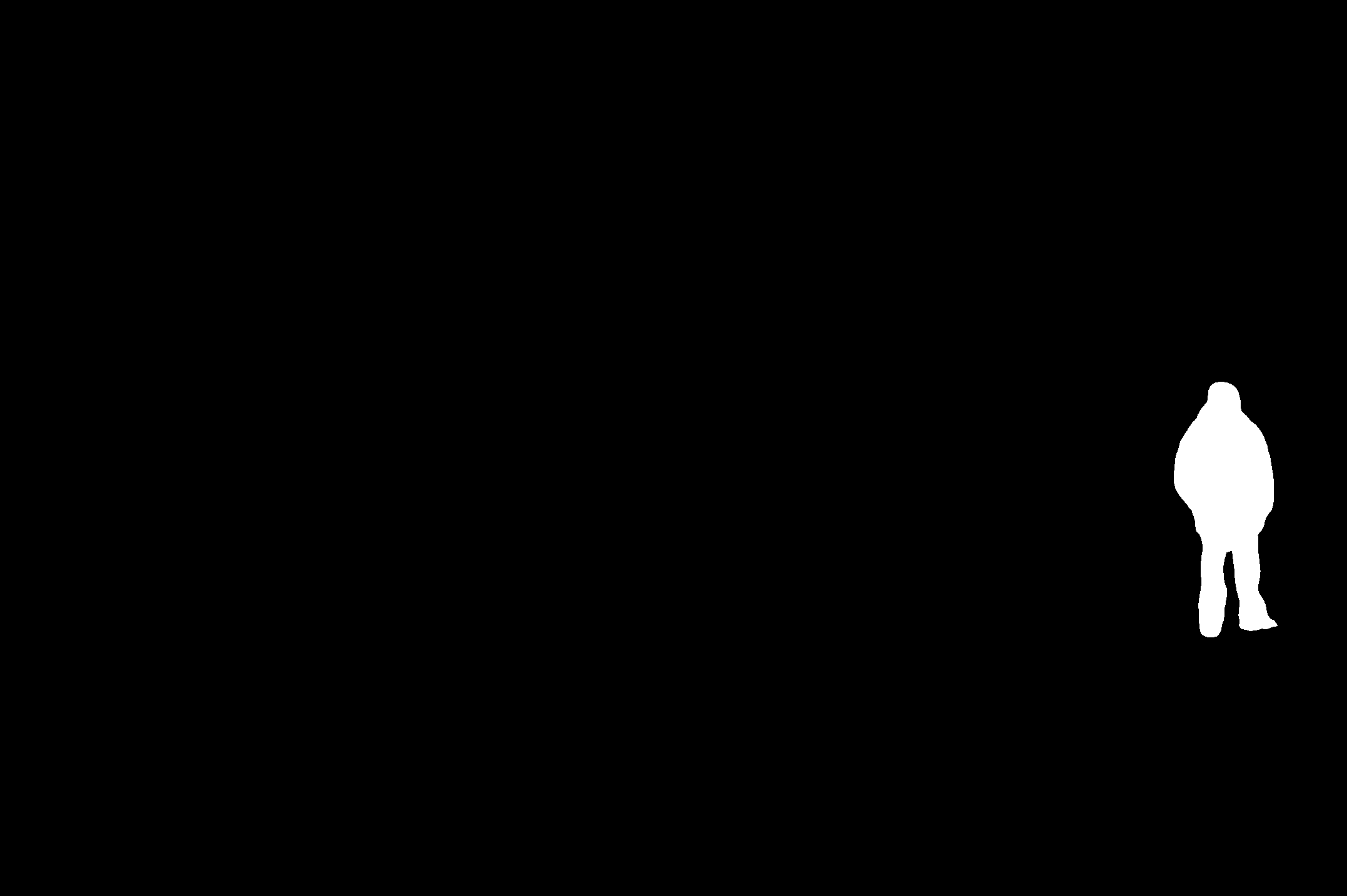}&%
\includegraphics[width=\s]{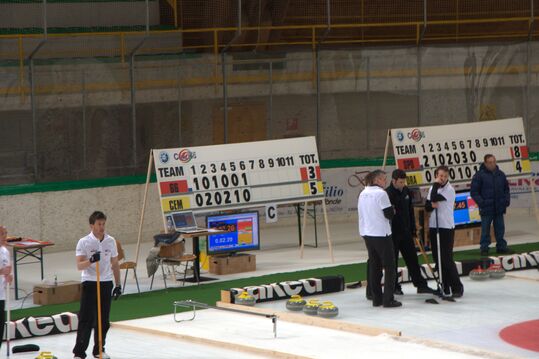}&%
\includegraphics[width=\s]{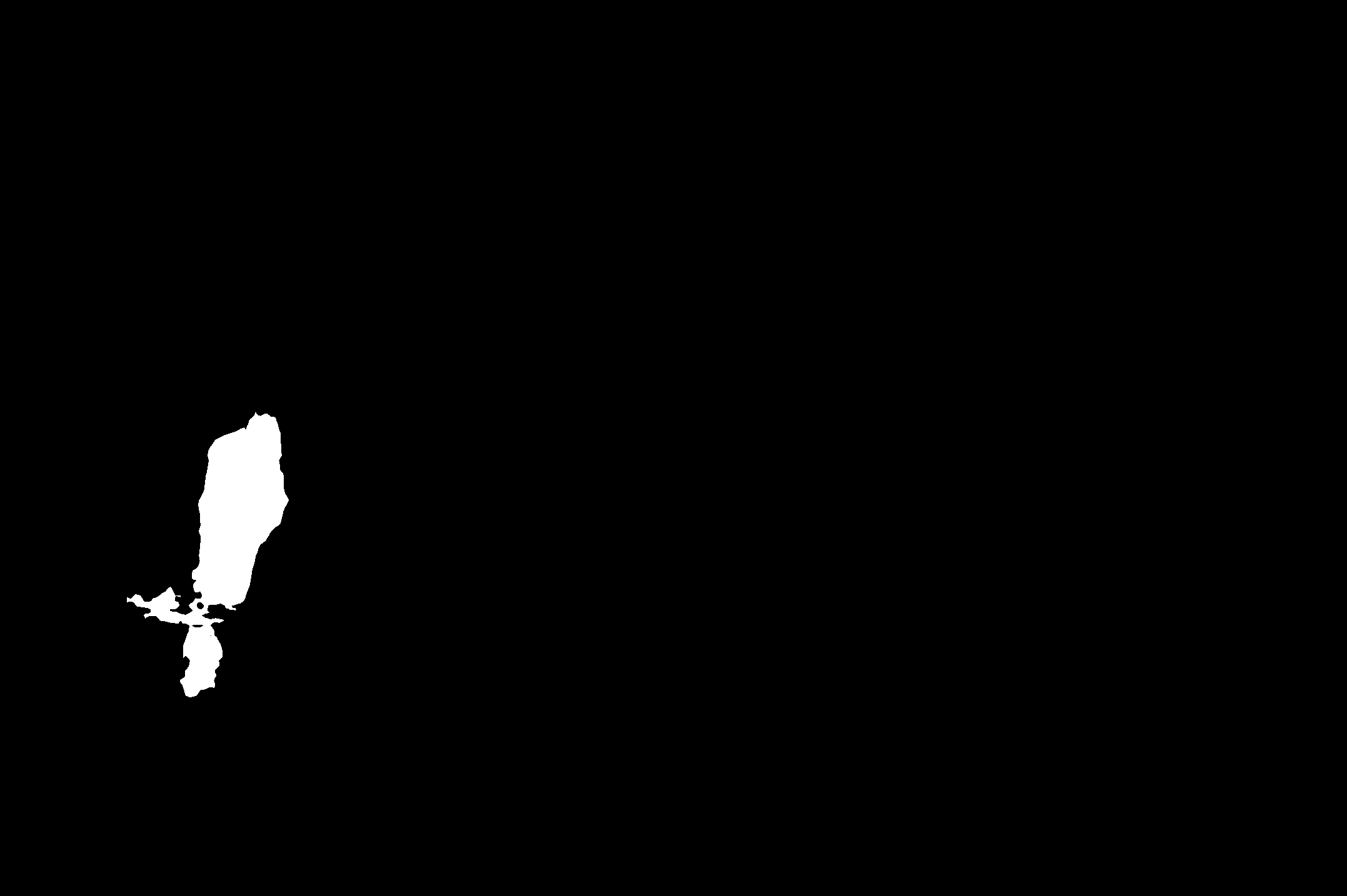}\\[\ss]
\includegraphics[width=\s]{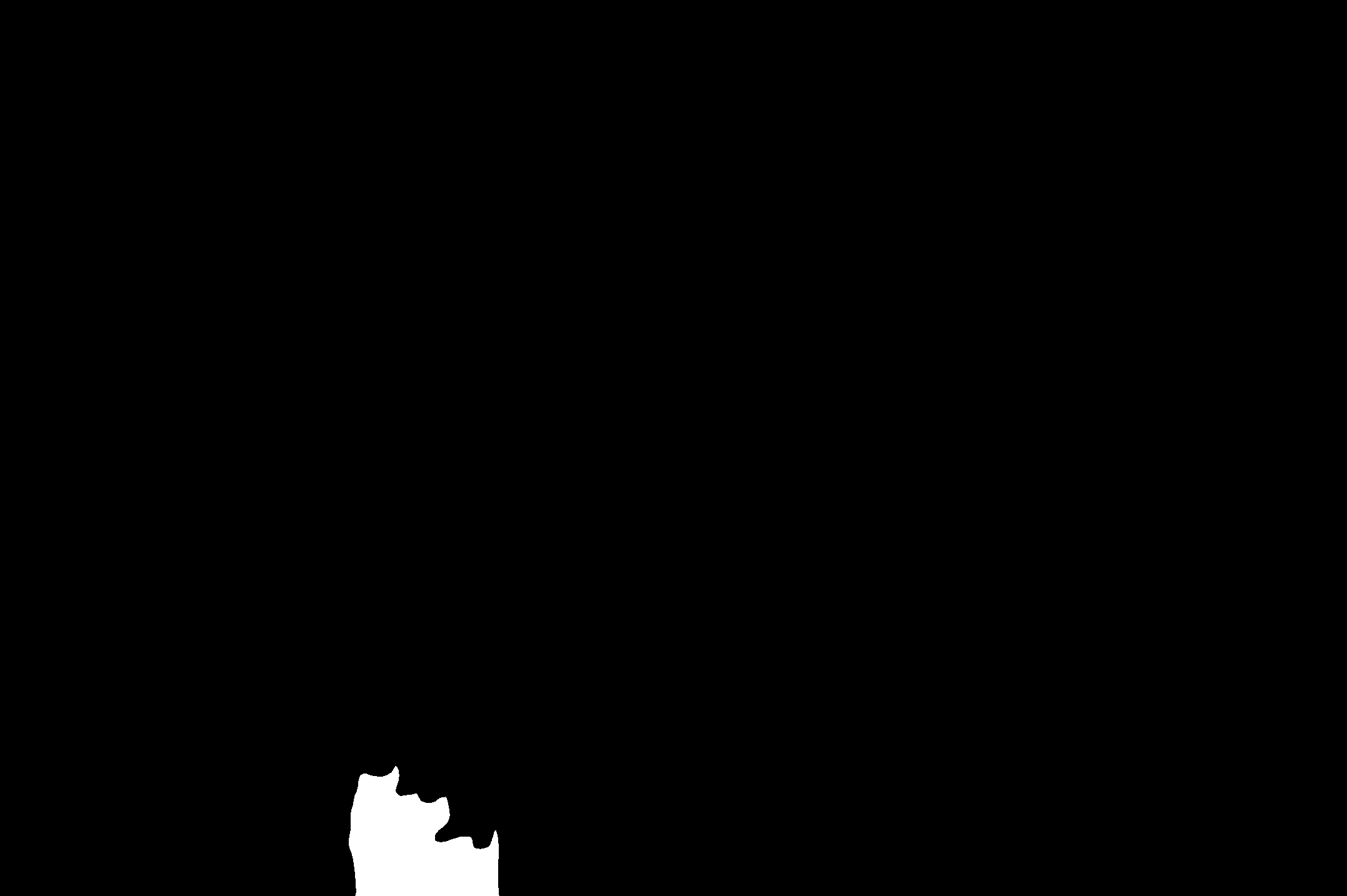}&%
\includegraphics[width=\s]{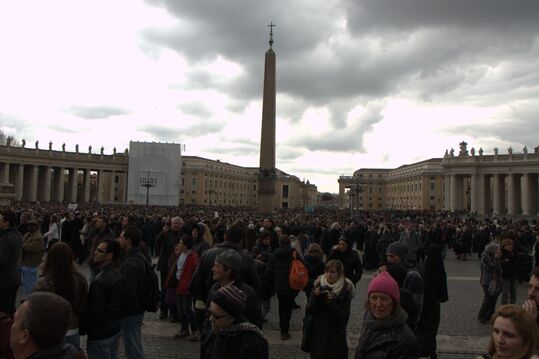}&%
\includegraphics[width=\s]{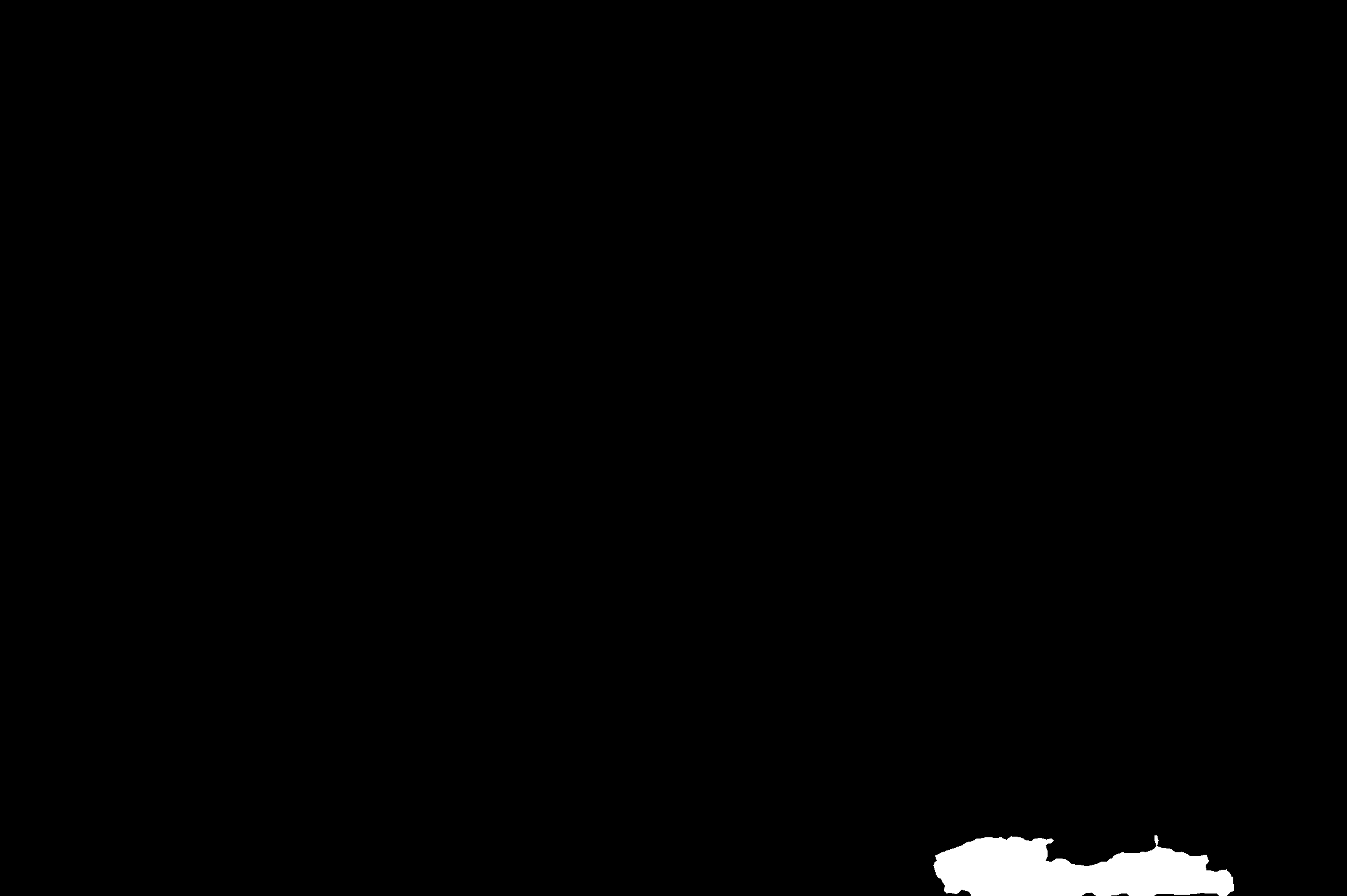}\\[\ss]
\includegraphics[width=\s]{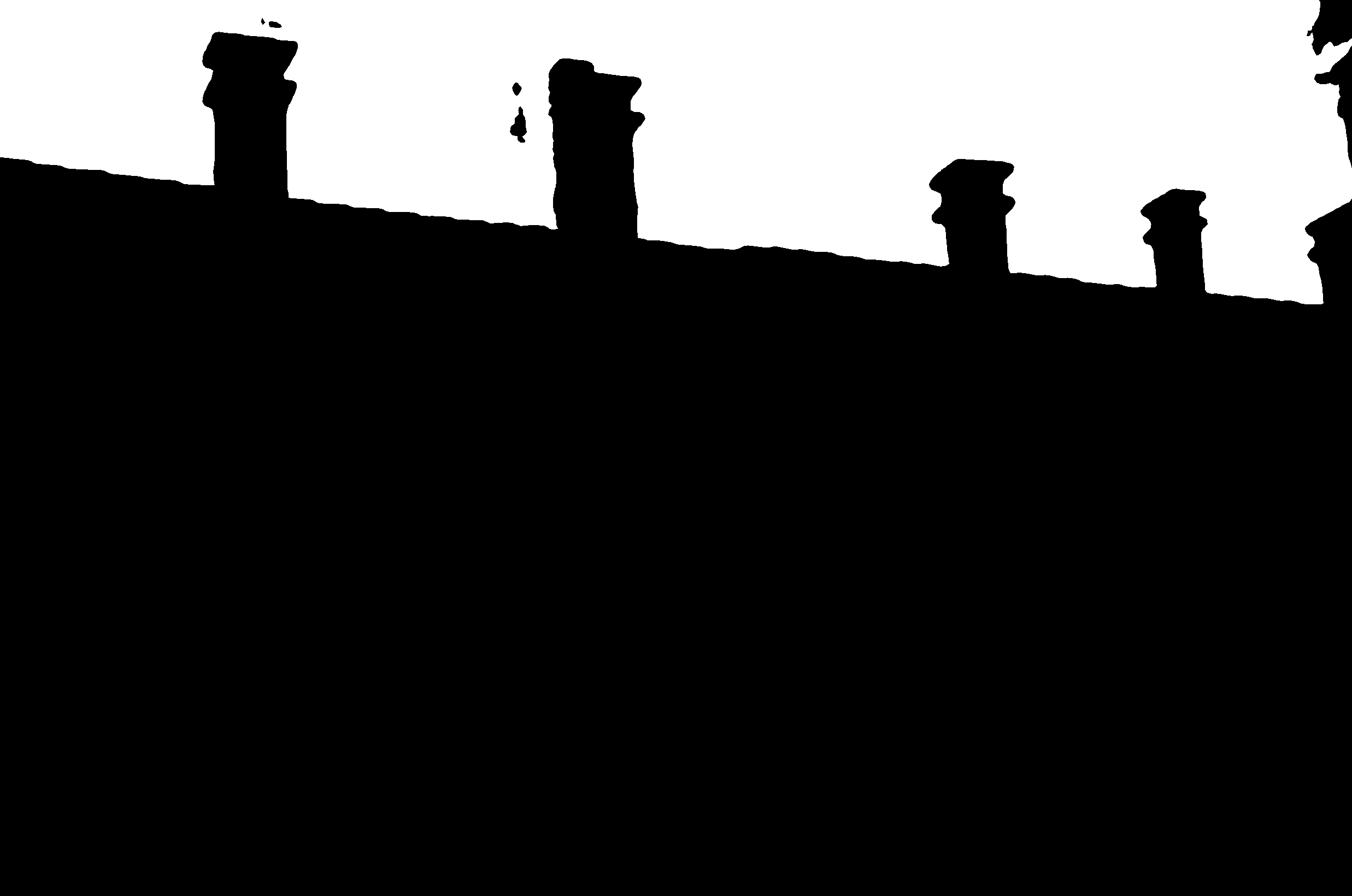}&%
\includegraphics[width=\s]{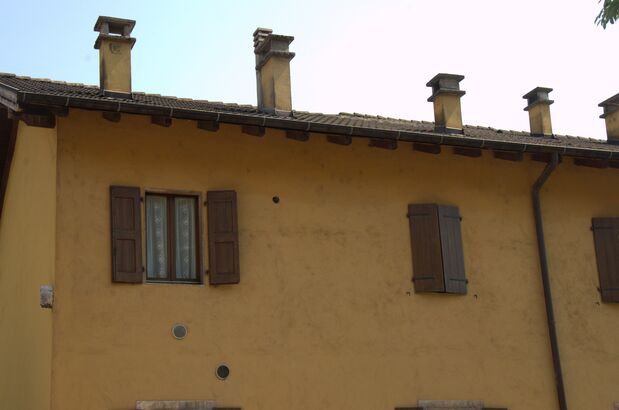}&%
\includegraphics[width=\s]{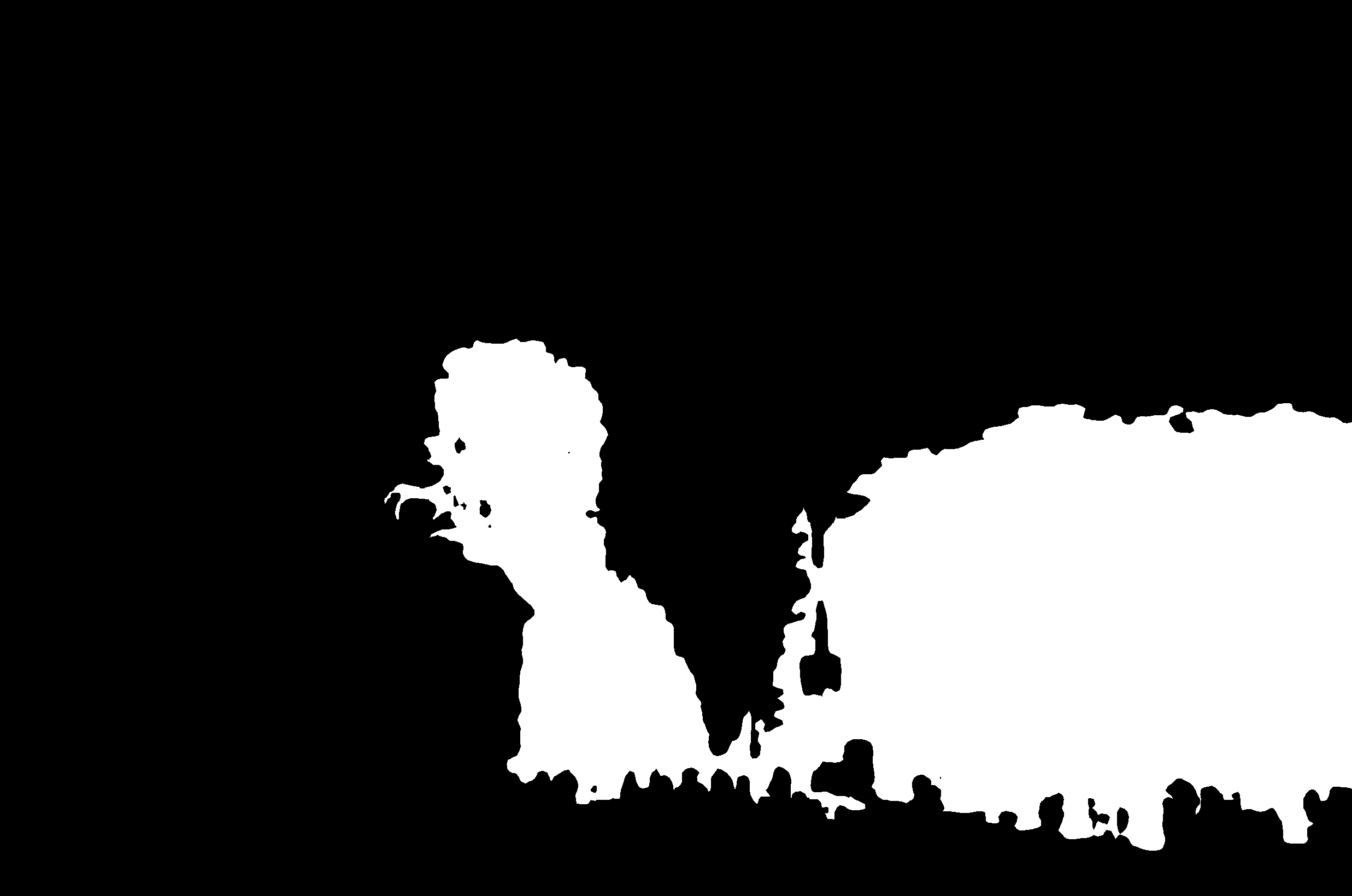}\\[\ss]
\includegraphics[width=\s]{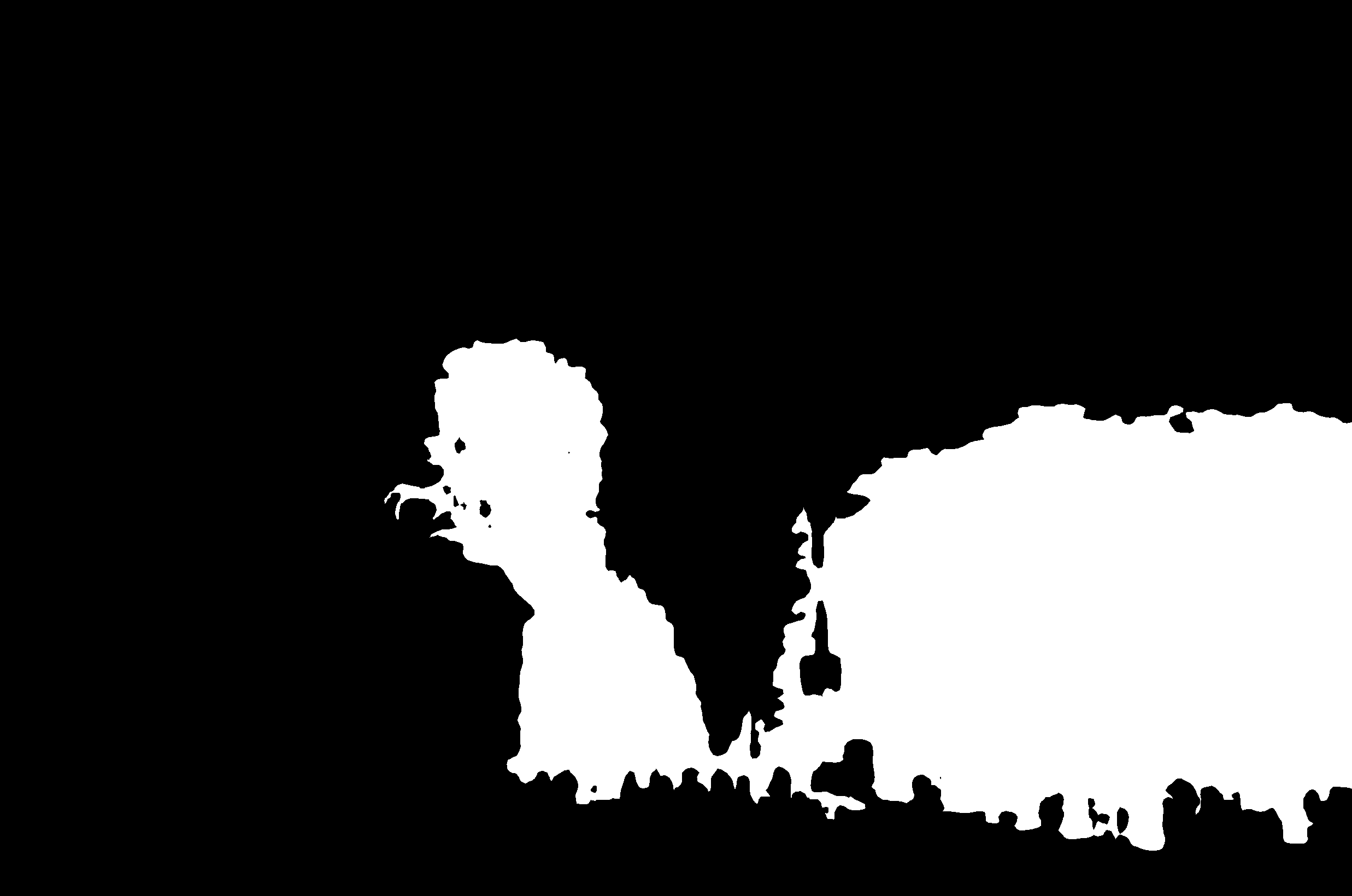}&%
\includegraphics[width=\s]{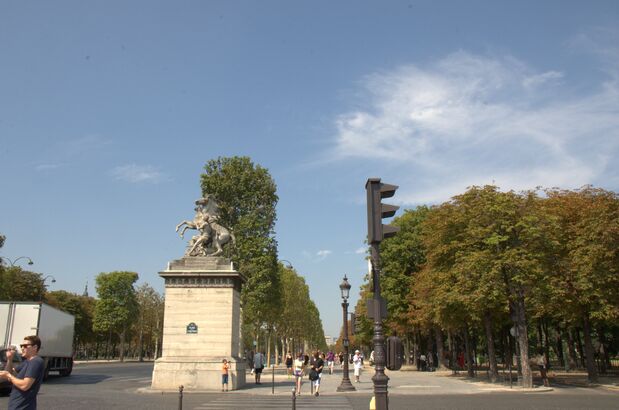}&%
\includegraphics[width=\s]{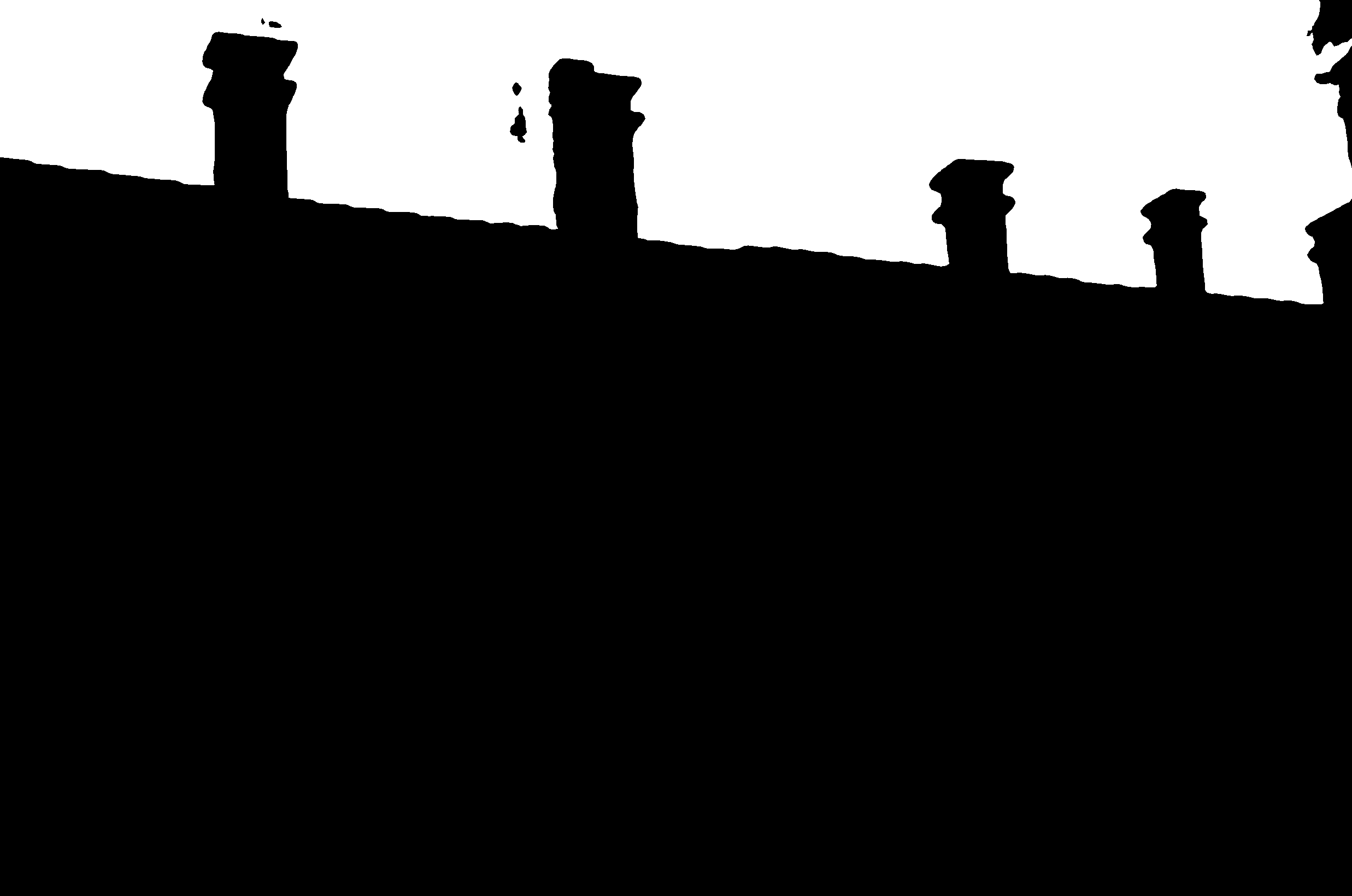}\\[\ss]
%\bottomrule
\end{tabular}
\caption{For each image, we use an endomask (left) taken from the image's segmentation, and an exomask (right) taken from another image and thus decorrelated from the image's contents. The last two images were paired during mask creation, thus the endomask of each becomes the exomask of the other.}
\label{fig:example_images}
\end{figure}

The second set of masks is unrelated to the image's content. We pair the images in order of their endomasks' sizes. The endomask of each image is then used as the exogenous mask, or \emph{exomask}, of its paired image. Using a mask from another image ensures the mask is not linked to the semantics of the image. The chosen pairing enables comparisons separately on each image, as the two masks of an image are of similar size. See Fig.~\ref{fig:example_images} for examples of endo- and exomasks.

\vspace{-10pt}
\paragraph{Multiple datasets}
%Not only do we want to check whether forensics tools are sensitive to inconsistencies in the image signature,
We want to know which inconsistencies, forensics tools are sensitive to. Changes in the image processing pipeline, done at different steps of the chain, lead to different inconsistencies (see Section~\ref{sec:pipeline}). As a consequence, we create five specific datasets, each of which feature a specific change in the image processing pipeline.

For each image, we start by randomly choosing the three parameters that are common to all datasets: 
\begin{itemize}
  \setlength\itemsep{0em}
    \item The mosaic pattern, chosen randomly among the four possible values, represents the horizontal and vertical offset of the first pixel sampled in red by the camera.
    \item The demosaicing algorithm, chosen randomly between bilinear filtering, VNG, PPG, AHD, DCB, DHT and AAHD from the LibRaw library~\cite{libraw}.
    \item The gamma-correction power, chosen uniformly between 1 and 2.5 (the typical range in most cameras).
\end{itemize}
Unless explicitly stated, those parameters stay the same between the two pipelines of the same image. For each image, both the endo- and exomasks, constructed as explained above, are the same across all datasets.

\vspace{-10pt}
\paragraph{Raw Noise Level dataset}
In this dataset we add random noise to the raw image before processing it. As pointed out in Section~\ref{sec:pipeline}, noise variance in raw images follows a linear relation given by $\sigma^2 = A + Bu$, where $A$ and $B$ are constants and $u$ is the noiseless image. We start by randomly selecting two different pairs of constants $(A_0, B_0)$ and $(A_1, B_1)$. $A_0$ and $A_1$ are sampled uniformly in $[0, 2]$, $B_0$ and $B_1$ in $[0, 6]$.
This range of values ensures that the resulting images look natural. 
To avoid cases in which both noise curves intersect, meaning  that there would be an intensity for which both images will have the same noise level, we impose a further condition: if $A_1$ is smaller (larger) than $A_0$ then $B_1$ must be also smaller (larger) than $B_0$. Finally, to obtain image $i$ ($i =0,1$), we add noise to the input image by randomly sampling, for each pixel $x$, a Gaussian distribution with 0 mean and $A_i + B_i u(x)$ variance.
%Finally, the image is obtained by randomly sampling, for each pixel $x$, a Gaussian distribution with 0 mean and $A_0 + B_0 u(x)$ variance; the same with variance $A_1 + B_1 u(x)$ for the second image.
Both images are then processed with the same pipeline. This dataset mimics the inconsistencies in noise models that could be found in spliced images.

\vspace{-10pt}
\paragraph{CFA Grid dataset}
In this dataset we only change the mosaic pattern of the forged image. Thus, the two images would be identical if not for their mosaic grids. This kind of trace could appear in the case of an internal copy-move. Indeed, even if the forged region has a similar signature, there is no reason the mosaic grid of the forged region should be the same as in the authentic region unless the copy-move translation is a multiple of 2 in both directions.

\vspace{-10pt}
\paragraph{CFA Algorithm dataset}
In this dataset, the two processing pipelines use different demosaicing algorithms. The demosaicing pattern is chosen independently for each pipeline, there is thus a $\frac 1 4$ chance that they are aligned.
%In this dataset the second processing pipeline uses a different demosaicing algorithm than the first.
A new mosaic pattern is also randomly chosen, thus having a $\frac 34$ chance of being different from the one of the main image.
This dataset represents the change in the mosaic that would occur from splicing, as two different images most likely do not share the same demosaicing algorithms, and the alignment of their patterns after splicing is random.%, as the large variety of existing demosaicing algorithms almost guarantee that the spliced region's algorithm will be different. 
%There is no reason why the spliced mosaic pattern should be aligned with the original.

\vspace{-10pt}
\paragraph{JPEG Grid dataset}

 In this dataset we only change the compression grid.
As with the CFA Grid dataset, in the case of an internal copy-move, the JPEG grid of the forged region does not need to be same as the one in the authentic image, unless the copy-move translation is a multiple of 8 in both directions. %As above, we force the grid to be different, even though this would only occur with a $\frac{63}{64}$ chance in practice. 
%PLAN Y: \qb{If we need space, we can combine either JPEG quality+CFA alg, and JPEG grid+CFA grid together (splicing/internal copymove), or JPEG quality+JPEG grid and CFA alg + CFA grid.} \tina{I would combine JPEG with JPEG and CFA with CFA.}
The JPEG compression quality is then chosen randomly between 75 and 100 (100 is the best quality), keeping the values in a range that is typical of most compressed images and challenging enough for JPEG-based algorithms.

\vspace{-10pt}
\paragraph{JPEG Quality dataset}
In this dataset both the authentic and forged regions are processed with the same pipeline, except for the JPEG compression which is done in the two regions with different quality factors, again chosen uniformly between 75 and 100. Like with the CFA Algorithm dataset, a new JPEG grid pattern is also randomly chosen, which has a $\frac{63}{64}$ chance of being different from the main region's grid. This represents 
%a change that would result from 
the splicing of an image onto another, both images being compressed at different quality factors.

\vspace{-10pt}
\paragraph{Hybrid dataset}
One could argue that although generic learning-based forensics tools may not be able to point out a single inconsistency in an image, they might be best suited to find multiple inconsistencies stacked together. Clearly a splicing may introduce joint inconsistencies in noise level, JPEG encoding and demosaicing; while a direct copy-move can introduce alterations in the JPEG and CFA grids.
To investigate such possibilities, in addition to the five specific datasets described above, we created a sixth, hybrid dataset. In this dataset, forgeries combine noise, demosaicing and/or JPEG compression traces. To create this dataset, we adopt the following procedure for each image:
\begin{enumerate}
\itemsep0em 
  \setlength\itemsep{0em}
    \item We randomly choose whether to modify two or three steps of the pipeline (added noise, demosaicing grid/method, JPEG grid/quality). If we only change two, we select which steps to change.
    \item For JPEG and CFA modifications, we select whether we only change the CFA and JPEG grids,
    %(like in the CFA grid and JPEG grid datasets)
    or if we change the demosaicing methods, the JPEG quality factor and potentially the CFA and JPEG grids.
    %(like in the CFA algorithm and JPEG quality datasets)
    The decision is made jointly for JPEG and CFA, as the CFA and JPEG Grid datasets mimic artefacts commonly found in internal copy-move forgeries, whereas the CFA Algorithm and JPEG Quality datasets represent inconsistencies more typical of splicing.
    \item Finally, for each different change, we select its parameters in the same way as for the specific datasets.
\end{enumerate}

\iffalse
\begin{itemize}
\item Take a pristine raw image
\item Process it in two ways and $I_0$ and $I_1$, with one element changing between the two ways (eg. grid of demosaicking, JPEG grid, double compression in one image, etc)
\item Segment $I_0$ (which algorithm to use for this?) and randomly take one region
\item In the selected region, substitute $I_1$ for $I_0$

\end{itemize}
\fi

\section{Experiments}\label{sec:experiments}

\subsection{Evaluated methods}

We use the constructed database to conduct an evaluation  of image forensics tools. We test both classic and state-of-the-art image forgery detection methods pertaining to different traces in an image: noise-based detection methods Splicebuster~\cite{splicebuster2015}, Lyu~\cite{lyupan2013, matlab} and Mahdian~\cite{mahdian2009, matlab}; CFA-grid detection methods Bammey~\cite{bammey}, Shin~\cite{shin2017color} and Choi~\cite{choi}; JPEG-based methods Zero~\cite{ZERO}, CAGI~\cite{CAGI, matlab}, FDF-A~\cite{ADQ3, matlab}, I-CDA~\cite{ADQ2, matlab}, CDA~\cite{ADQ1, matlab} and BAG~\cite{BLK, matlab}, as well as neural-network-based generic methods Noiseprint~\cite{noiseprint}, ManTraNet~\cite{mantranet} and Self-Consistency~\cite{selfconsistency}.

\subsection{Evaluation Metrics}
\label{sec:metrics}

We evaluate the results of these methods using the Matthews correlation coefficient (MCC)~\cite{MATTHEWS1975442}. This metric varies from -1 for a detection that is complementary to the ground truth, to 1 for a perfect detection. A score of 0 represents an uninformative result and is the expected performance of any random classifier. The MCC is more representative than the F1 and IoU scores~\cite{mcc1, mcc2}, partly as it is less dependant on the proportion of positives in the ground truth, which is especially important given the large variety of forgery mask sizes in the database.
%It is defined as \qb{probably delete the formula: standard measure and takes a lot of space}
%\[\scriptstyle \phantom{.}MCC = \frac{TP\times TN - FP\times FN}{\sqrt{(TP+FP)\cdot(TP+FN)\cdot(TN+FP)\cdot(TN+FN)}}.\]
The MCC is computed for each image, and then averaged over each dataset.
As most surveyed methods do not provide a binary output but a continuous heatmap, we weight the confusion matrix using the heatmap. See the supplementary materials for more details, as well as for score tables with other metrics.

\subsection{Results}

\begin{figure*}[ht!]
\centering{
\def\s{0.145\textwidth}
\setlength{\tabcolsep}{0.001em}
\begin{tabular}{lcccccc}
%\toprule
&Noise Level&CFA Grid&CFA Algorithm&JPEG Grid&JPEG Quality&Hybrid\\\raisebox{16pt}{Splicebuster~\cite{splicebuster2015}}%
&\includegraphics[width=\s]{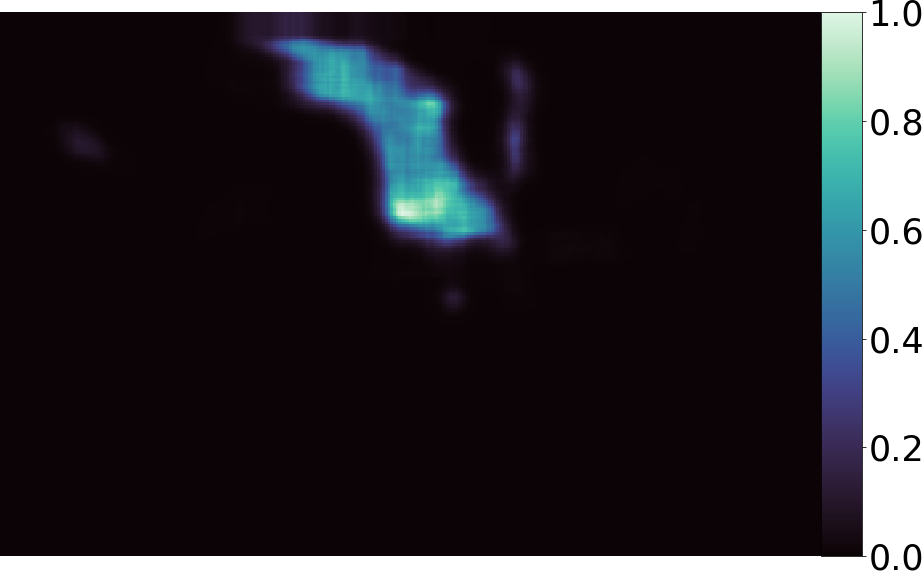}%
&\includegraphics[width=\s]{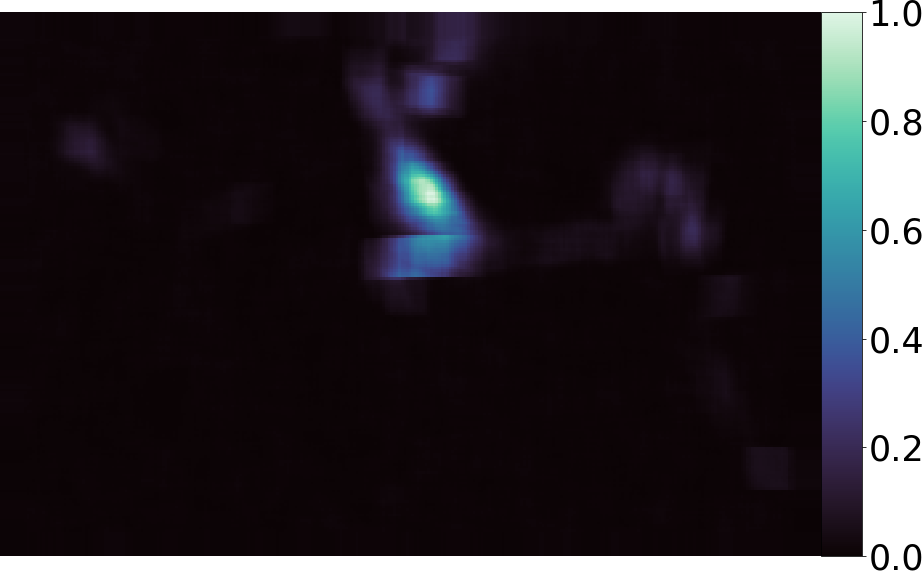}%
&\includegraphics[width=\s]{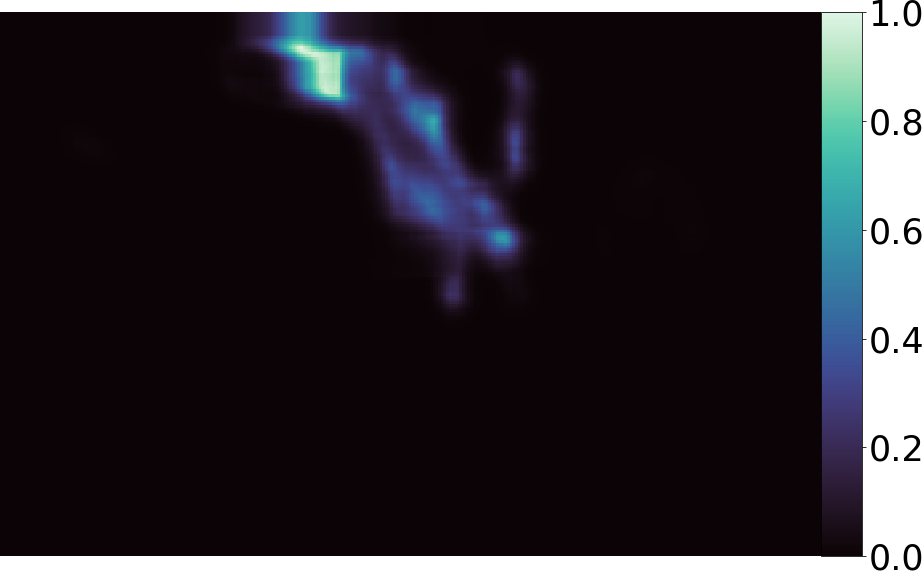}%
&\includegraphics[width=\s]{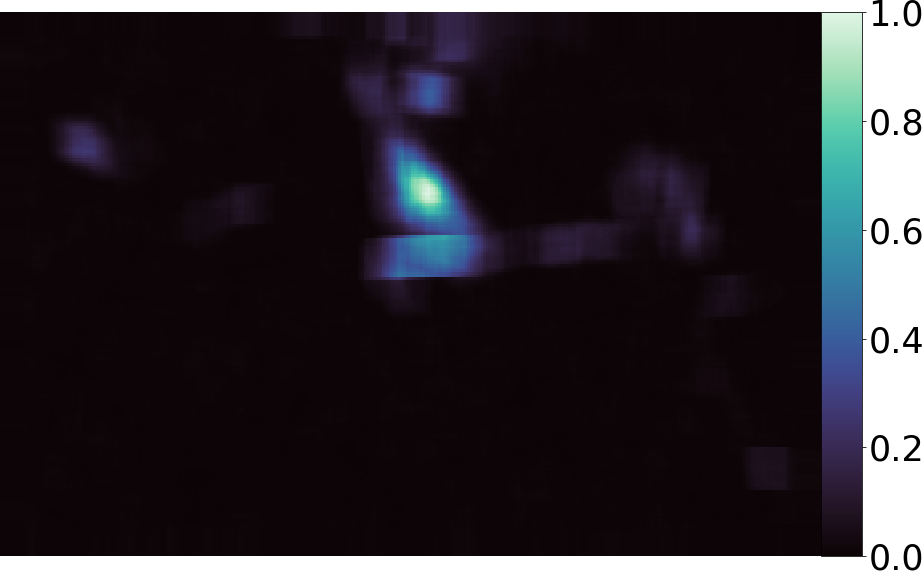}%
&\includegraphics[width=\s]{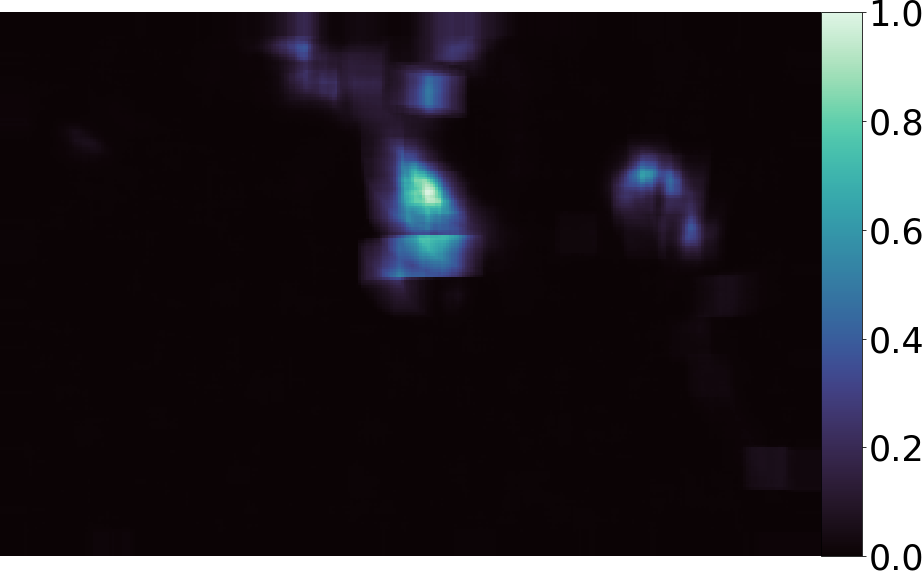}%
&\includegraphics[width=\s]{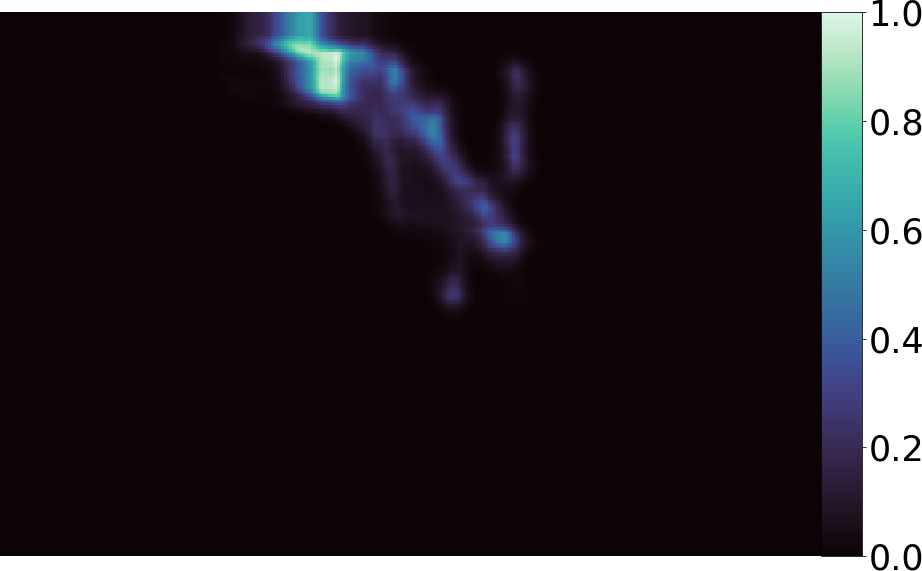}\\
\raisebox{16pt}{Bammey~\cite{bammey}}%
&\includegraphics[width=\s]{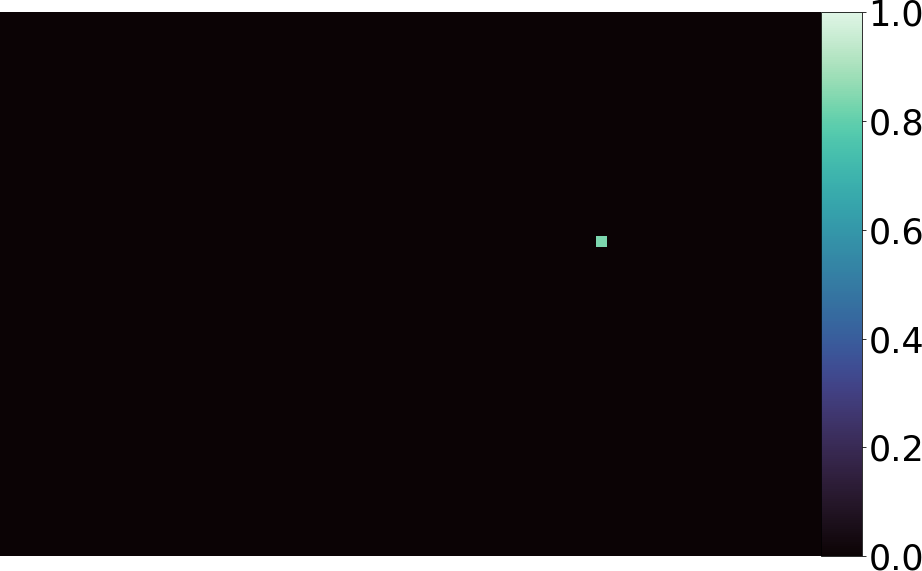}%
&\includegraphics[width=\s]{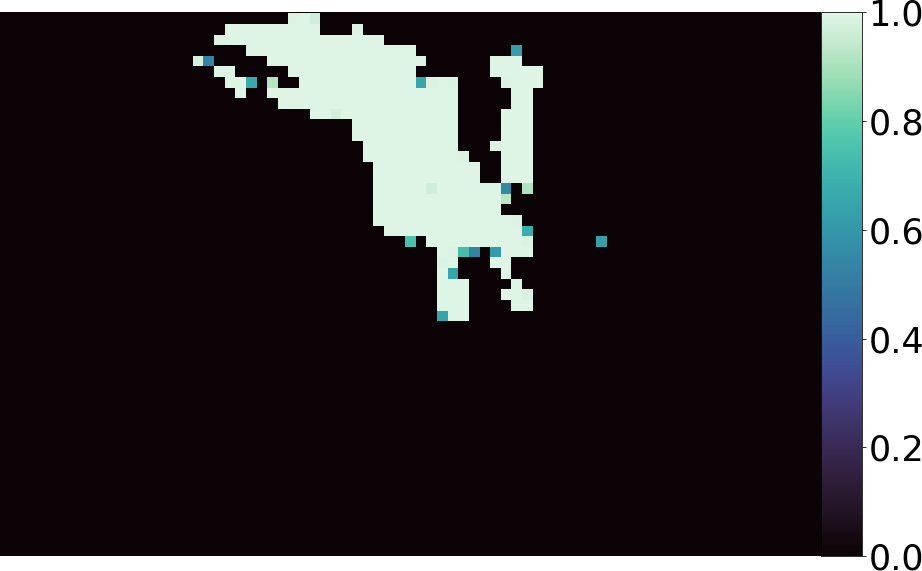}%
&\includegraphics[width=\s]{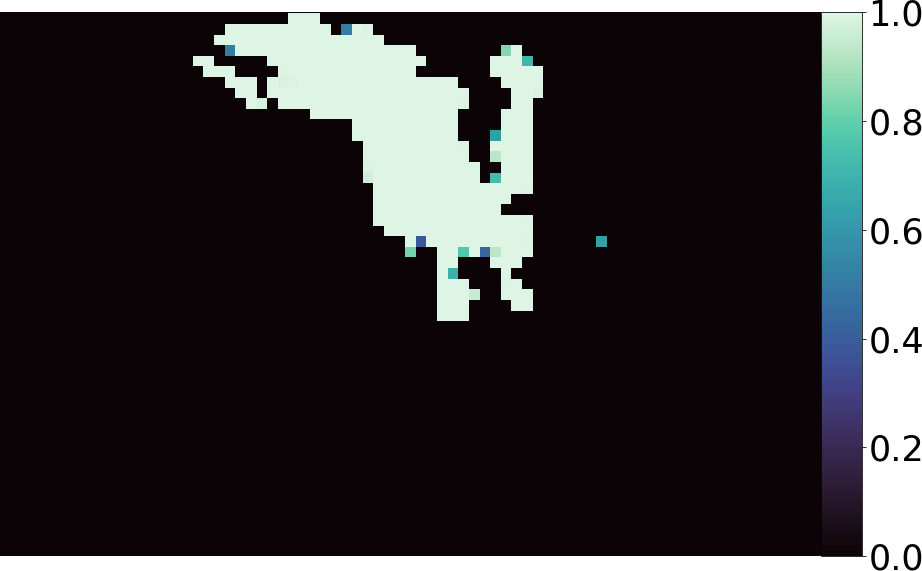}%
&\includegraphics[width=\s]{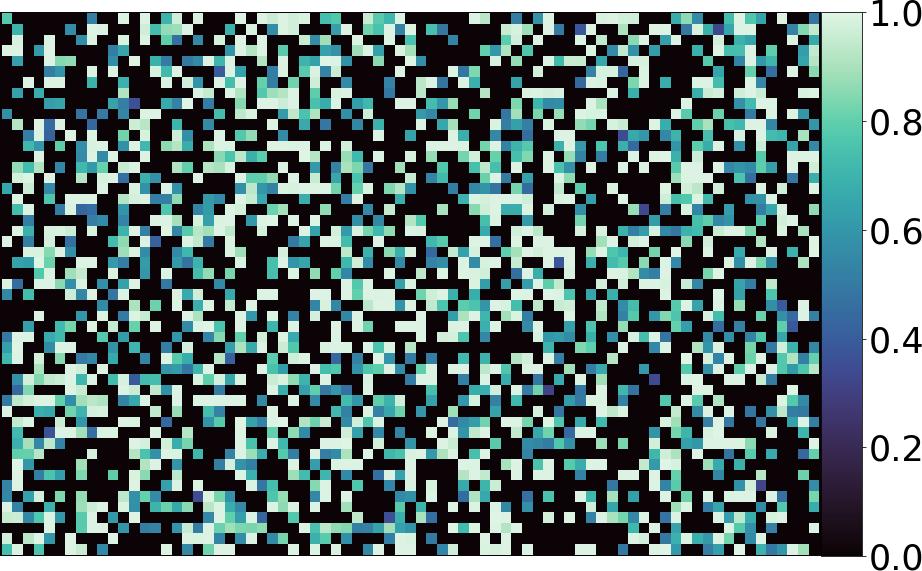}%
&\includegraphics[width=\s]{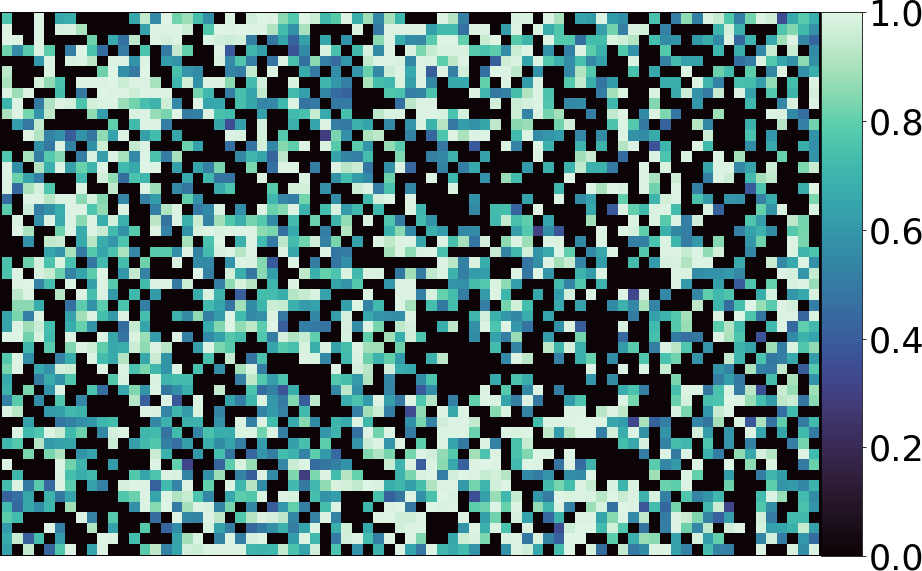}%
&\includegraphics[width=\s]{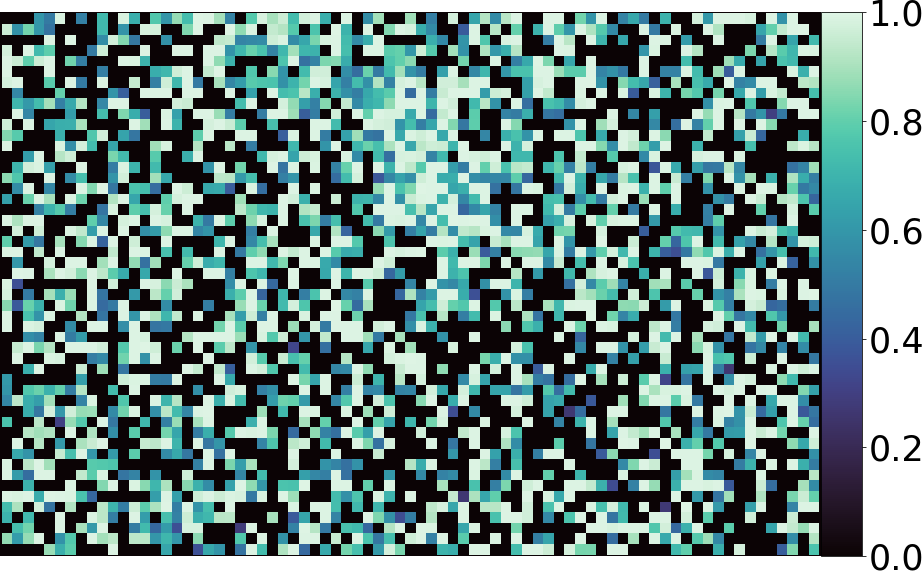}\\
\raisebox{16pt}{ZERO~\cite{ZERO}}%
&\includegraphics[width=\s]{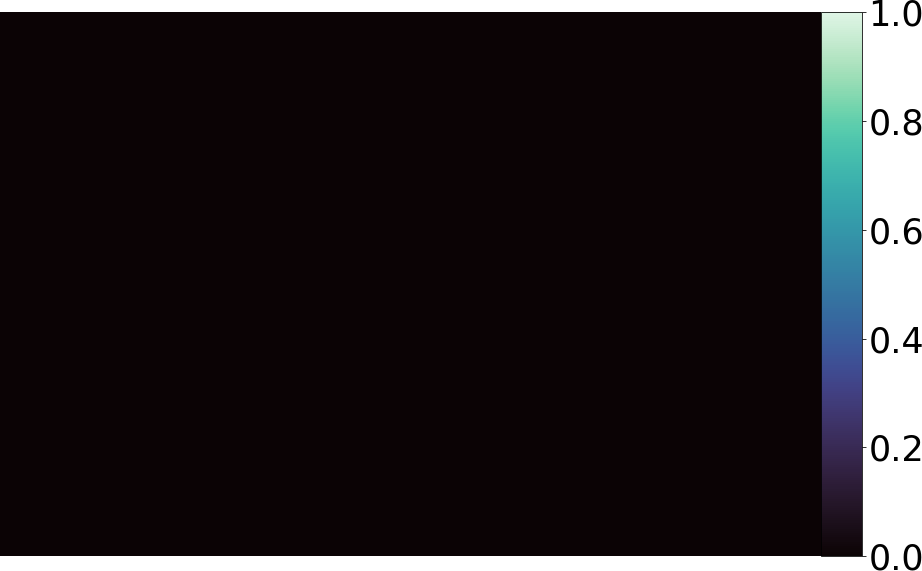}%
&\includegraphics[width=\s]{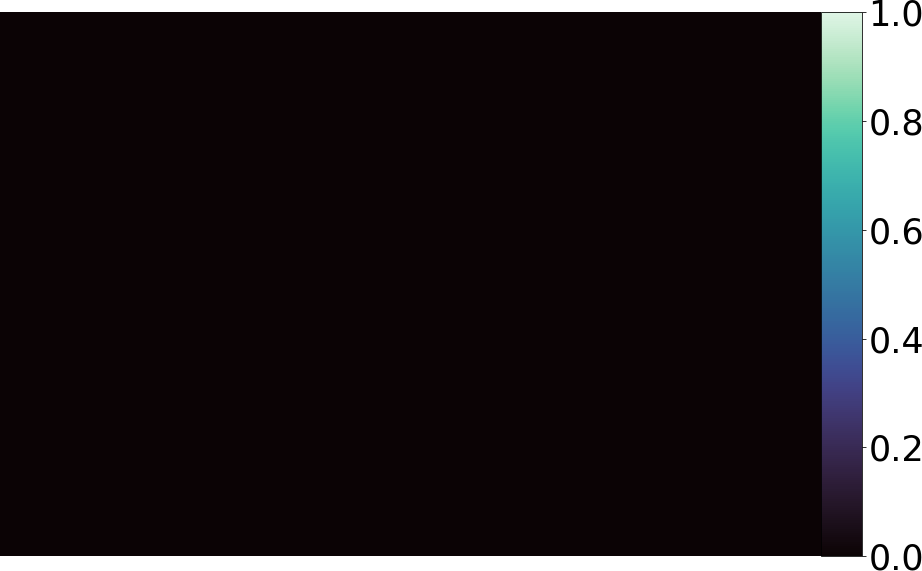}%
&\includegraphics[width=\s]{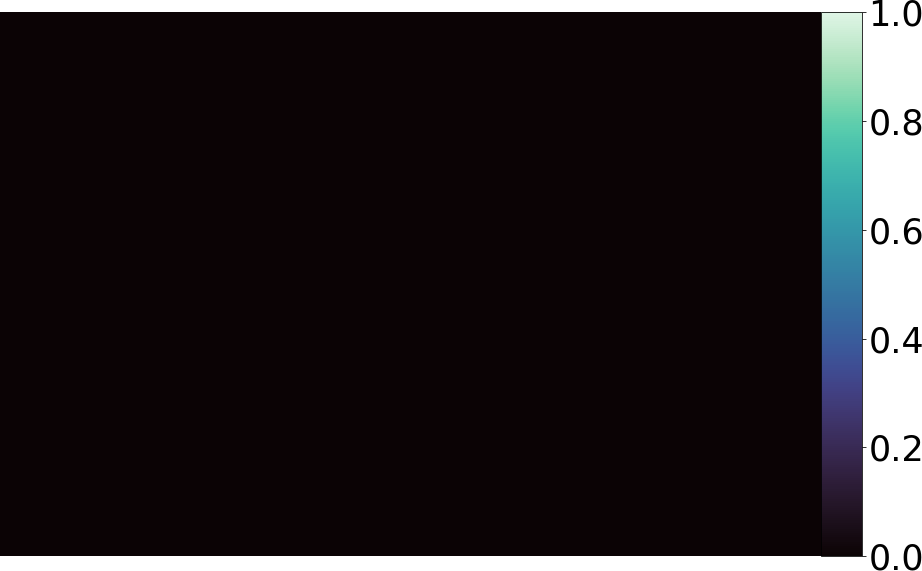}%
&\includegraphics[width=\s]{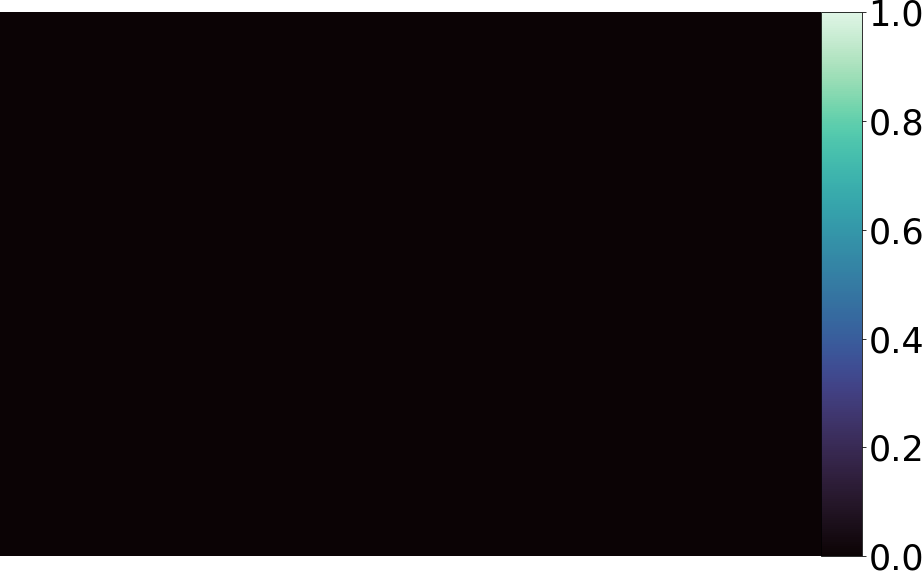}%
&\includegraphics[width=\s]{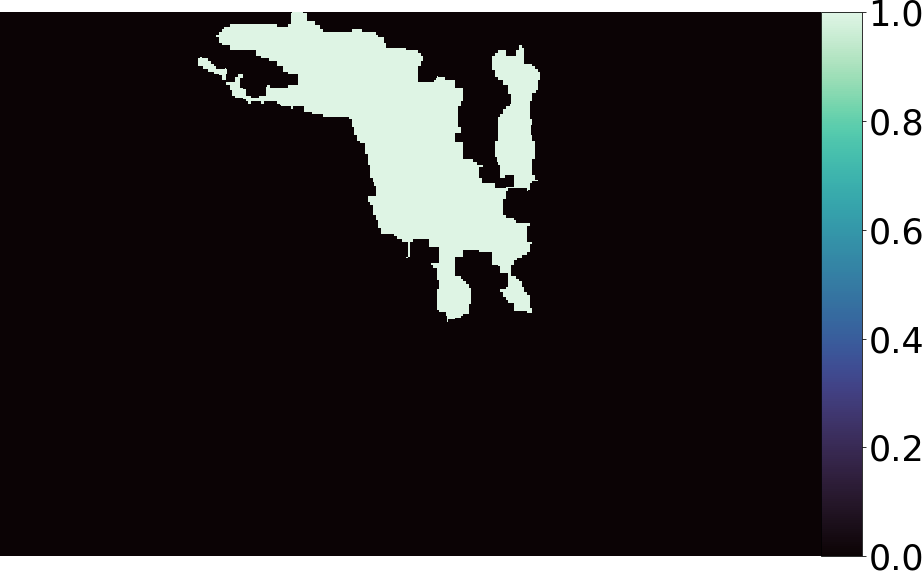}%
&\includegraphics[width=\s]{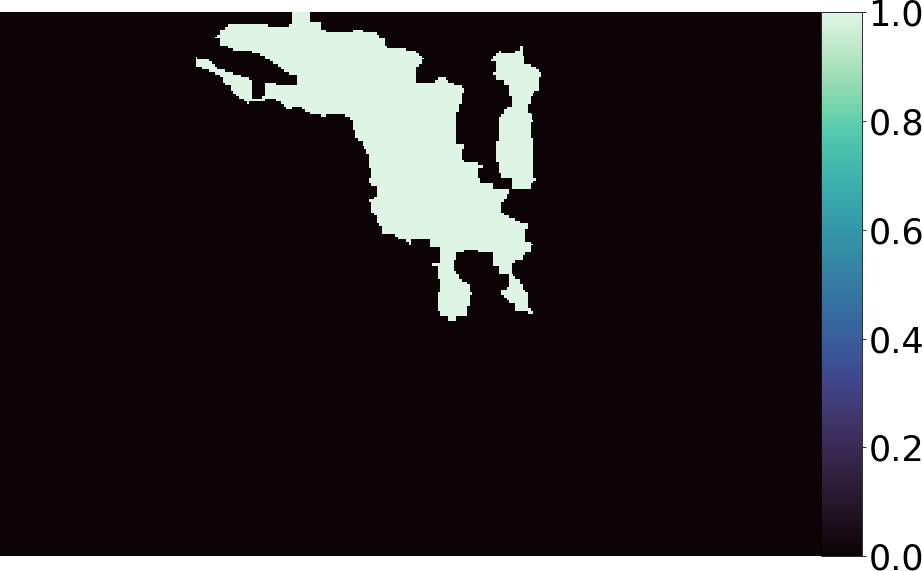}\\
\raisebox{16pt}{Noiseprint~\cite{noiseprint}}%
&\includegraphics[width=\s]{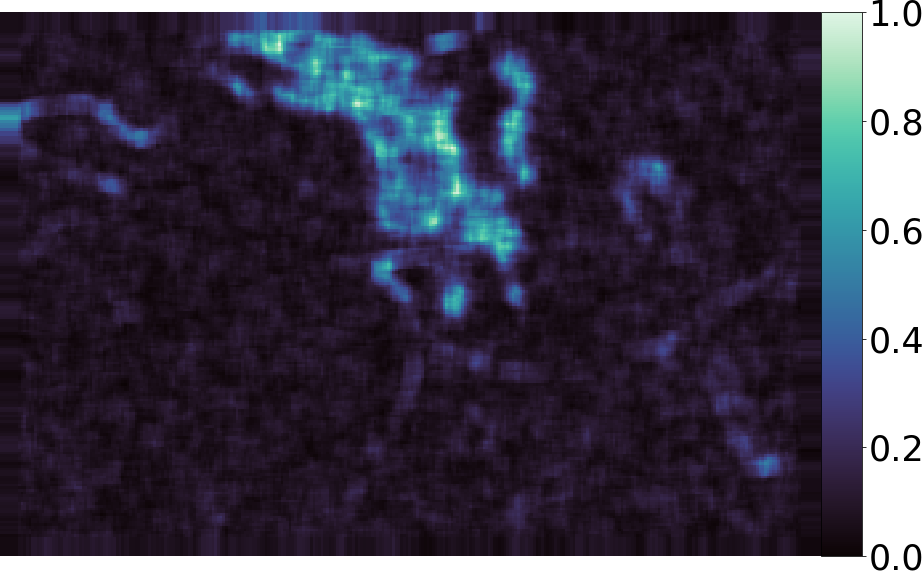}%
&\includegraphics[width=\s]{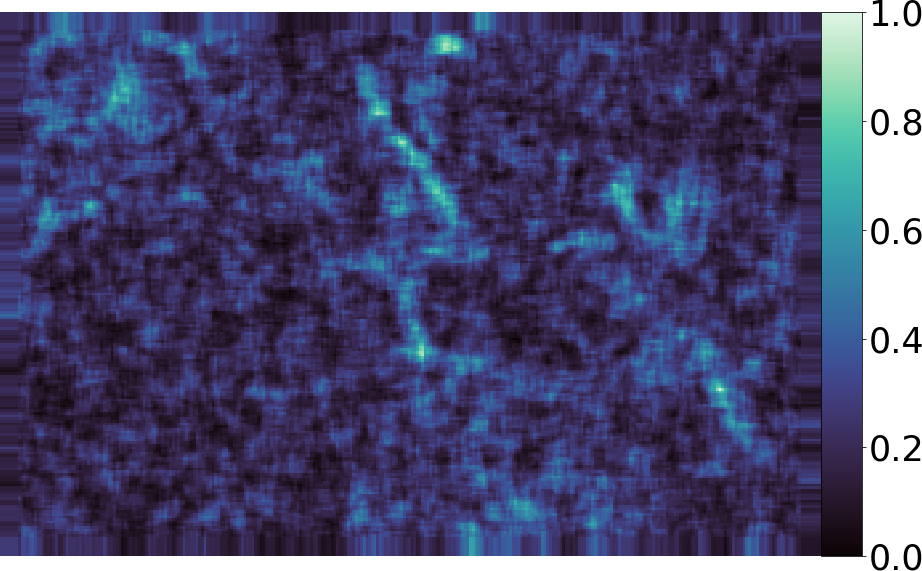}%
&\includegraphics[width=\s]{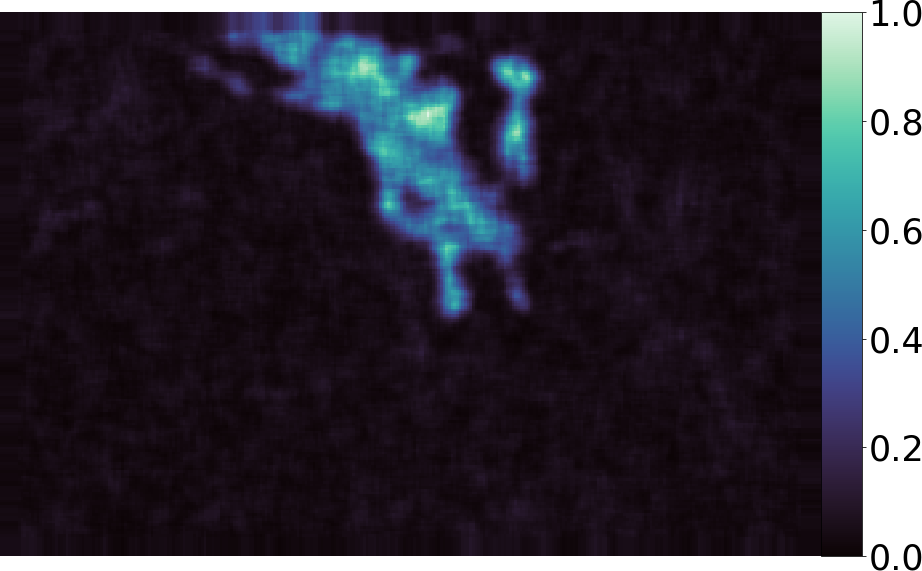}%
&\includegraphics[width=\s]{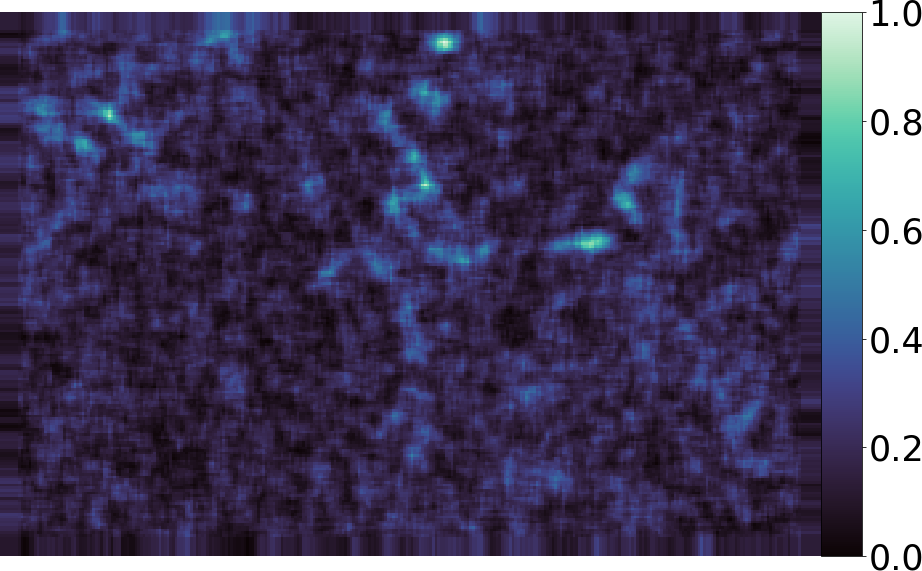}%
&\includegraphics[width=\s]{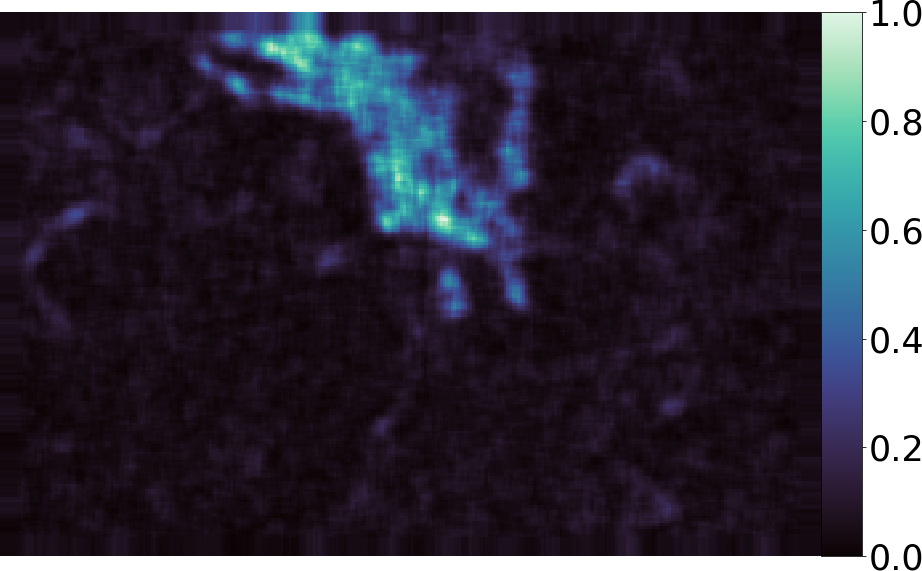}%
&\includegraphics[width=\s]{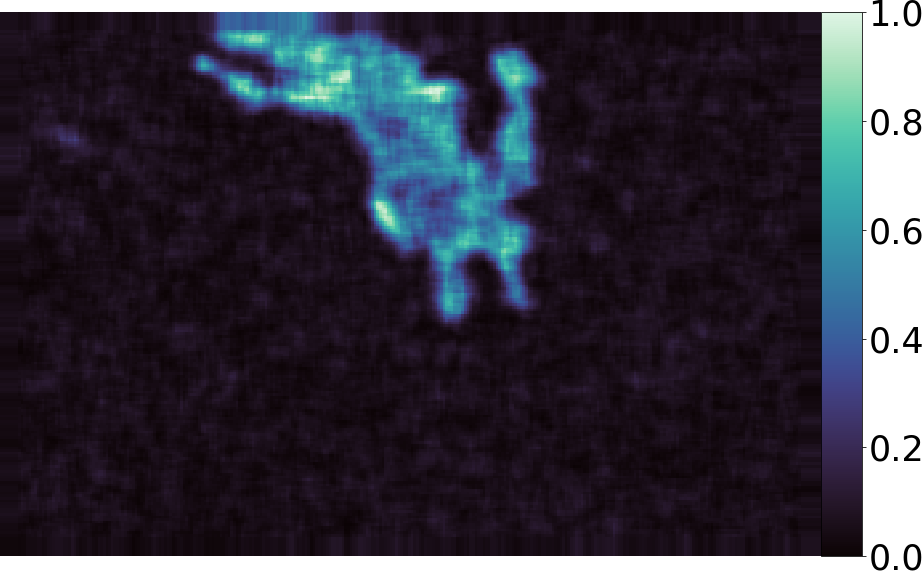}\\
\raisebox{16pt}{ManTraNet~\cite{mantranet}}%
&\includegraphics[width=\s]{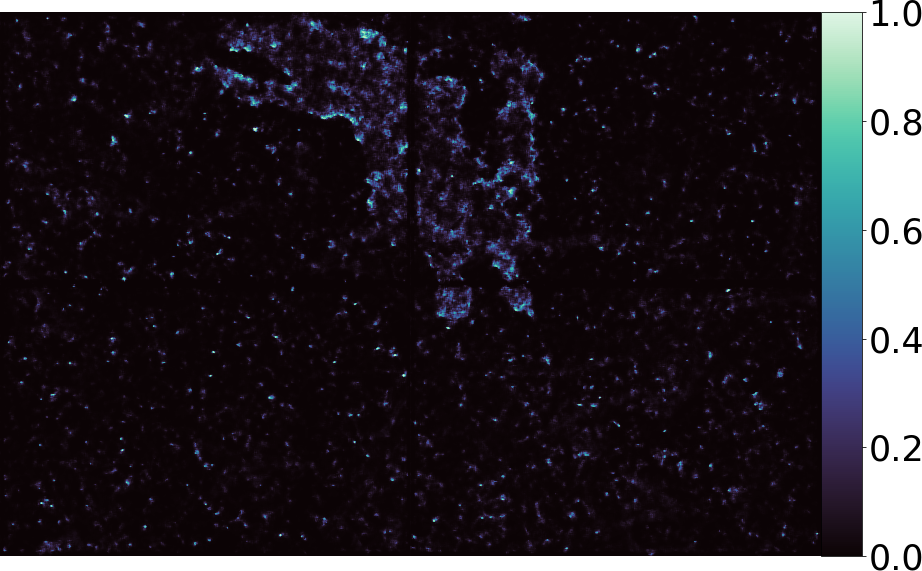}%
&\includegraphics[width=\s]{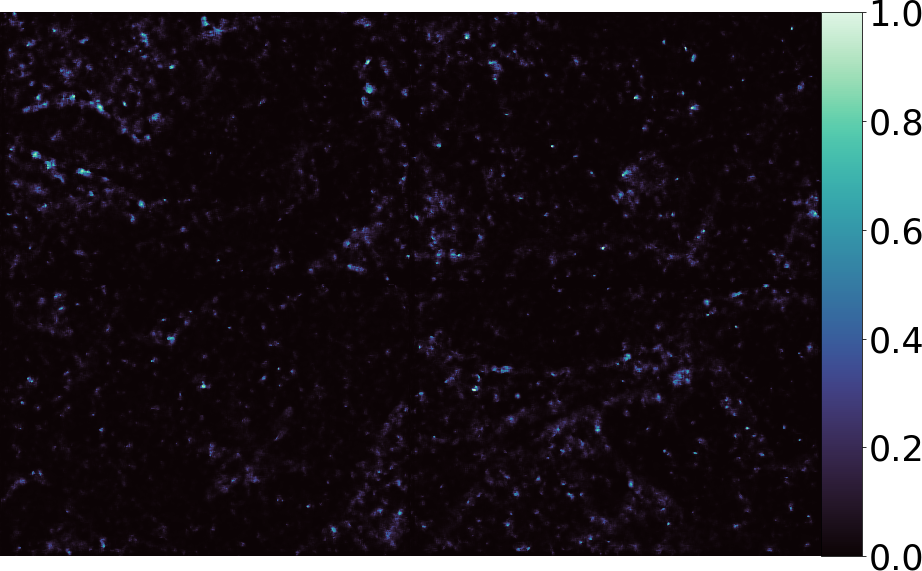}%
&\includegraphics[width=\s]{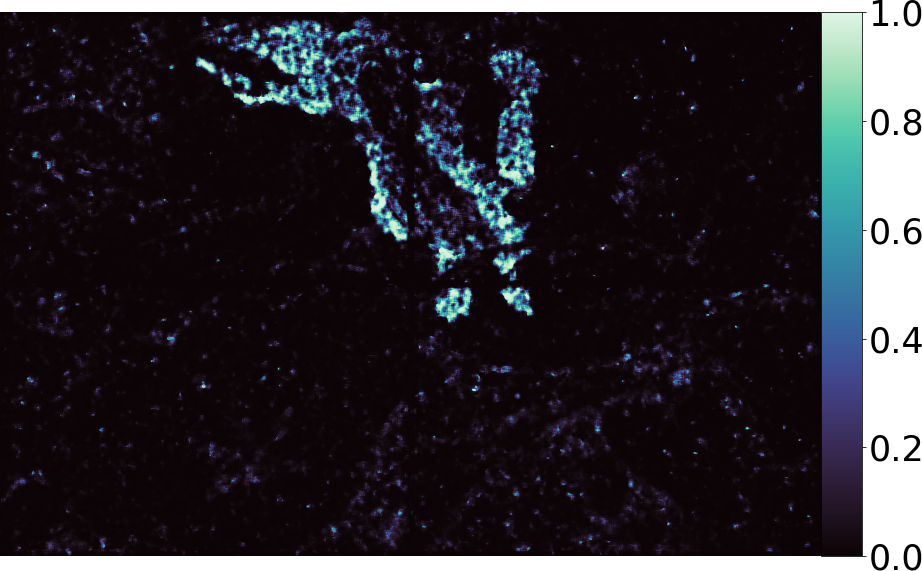}%
&\includegraphics[width=\s]{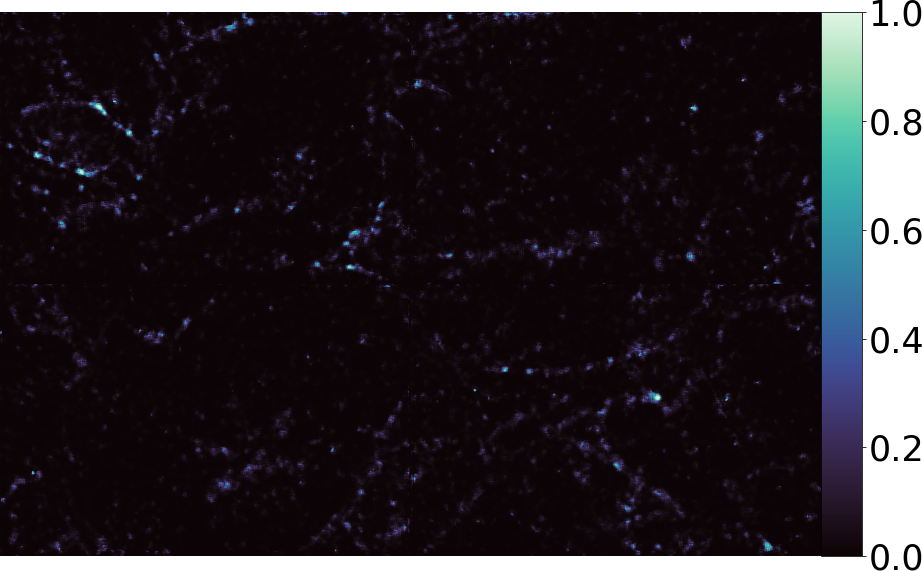}%
&\includegraphics[width=\s]{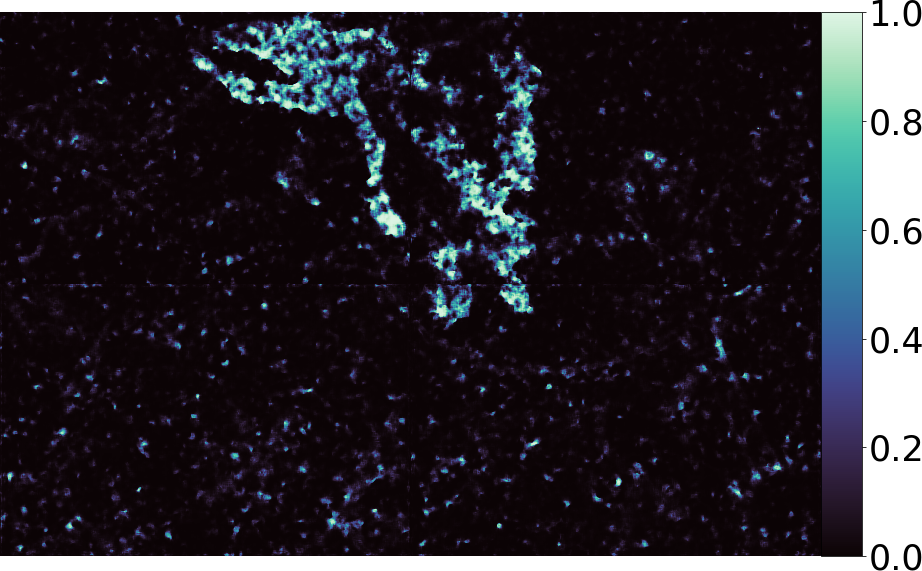}%
&\includegraphics[width=\s]{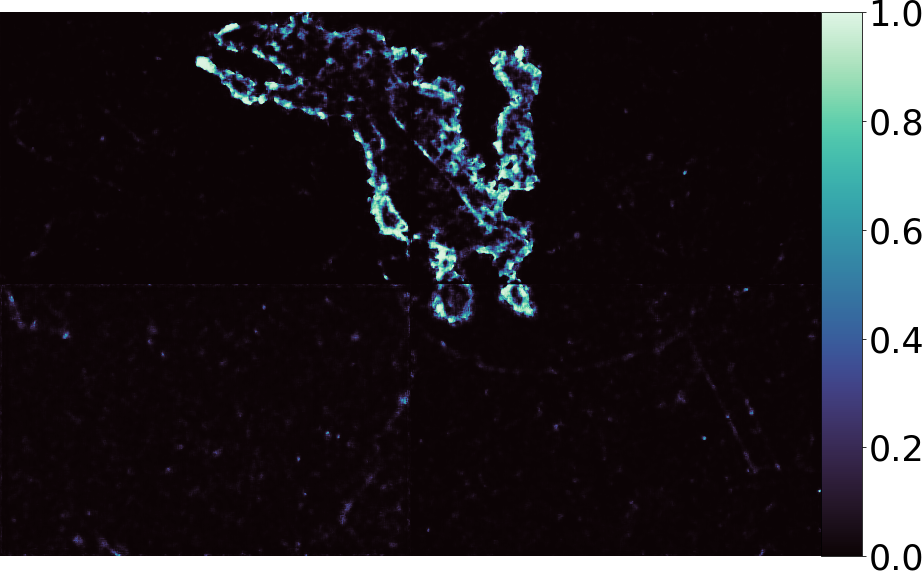}\\
\end{tabular}
}
\caption{Visualization of the results of several methods for one image on all the datasets. Some methods, such as Noiseprint or Bammey, correctly detect the forgeries in the relevant images, but tend to make noise-like false detections in the images for which they cannot see the forgery. Automatically selecting the relevant detections of an algorithm would make it easier to use without needing interpretation. The image and mask can be seen in Fig.~\ref{teaser}.}
\label{fig:experiment_buste}
\vspace{-10pt}
\end{figure*}

The complete results are given in Table \ref{fig:table}. Visualization of the detection by several methods on one image across all datasets can be seen in Figure~\ref{fig:experiment_buste}.
In the CFA and JPEG datasets, state-of-the-art methods that focus on those specific artefacts, such as Bammey~\cite{bammey} for CFA and ZERO~\cite{ZERO} for JPEG, perform much better than generic, neural-network-based tools. This is partly expected, as those methods aim to detect exactly this type of traces. This observation is more nuanced in the Noise Level dataset, where both %the noise-specific
Splicebuster~\cite{splicebuster2015} and % the more generic
Noiseprint~\cite{noiseprint} work equally well.%\qb{This last sentence should be discussed, since we're not sure how to classify splicebuster, and noiseprint has noise in its name (no matter what we say to justify it as generic)} \marina{I would say that Splicebuster is noise-based rather than noise-specific: it uses noise residual as an input but by computing the co-occurences of noise it is able to spot other traces in all of the steps that affect noise, not only the raw noise levels differences. On the other hand, Noiseprint uses as an input the fingerprint obtained as previously explained which is mainly noisy but that also contains other traces specific to the camera.}

\definecolor{c0}{HTML}{0E918C}
\definecolor{c1}{HTML}{F6830F}
%\definecolor{c3}{HTML}{1F3C88}
\definecolor{c3}{HTML}{1F3C88}

\definecolor{c2}{HTML}{BB2205}

\definecolor{grayA}{HTML}{D2D3C9}
\definecolor{cgray}{HTML}{708090}
\definecolor{cgreen}{HTML}{1F8C88}
\newcommand{\bIoU}[1]{\textcolor{c0}{#1}}
\newcommand{\sIoU}[1]{\textcolor{c1}{#1}}
\newcommand{\gIoU}[1]{\textcolor{c2}{#1}}
\newcommand{\lIoU}[1]{\textcolor{c3}{#1}}
\contourlength{0.15pt}
\contournumber{10}%
\newcommand{\ASEMREL}[2]{{\contour{c1}{\textcolor{c1}{#1~\footnotesize{(#2)}}}}}
\newcommand{\SEMREL}[2]{\contour{c3}{\textcolor{c3}{#1~\footnotesize{(#2)}}}}
\newcommand{\ASEMIR}[2]{\textcolor{c1!35!grayA}{#1~\footnotesize{(#2)}}}
\newcommand{\SEMIR}[2]{\textcolor{c3!35!grayA}{#1~\footnotesize{(#2)}}}
\newcommand{\BEST}[1]{\underline{#1}}

\newcommand{\NV}{\textcolor{grayA}{\hspace{6pt}---\hspace{6pt}}}

\newcommand{\IRRELEVANT}[1]{#1}
\newcommand{\RELEVANT}[1]{\cellcolor{cgreen!25}#1}
\newcommand{\tableScoreSeparator}{\hspace{4pt} }
\newcommand{\tableVerticalSeparator}{\vspace{4pt}}
\newcommand{\TABLETBD}{\textcolor{red}{\textbf{TBD}}}
\newcommand{\tbest}[1]{\textbf{#1}}

\begin{table*}[ht!]
\centering{
\begin{tabular}{llcccccc}
\toprule
&&&&Dataset\vspace{7pt}\\
&&Noise Level&CFA Grid&CFA Algorithm&JPEG Grid&JPEG Quality&Hybrid\\\midrule\multirow{6}{*}{\rotatebox{90}{Noise-based}}&\multirow{2}{*}{Splicebuster~\cite{splicebuster2015}}&\ASEMREL{0.099}{0.188}&\ASEMIR{0.003}{0.085}&\ASEMREL{0.075}{0.157}&\ASEMIR{0.005}{0.083}&\ASEMREL{0.084}{0.175}&\ASEMREL{0.101}{0.192}\\
&&\SEMREL{0.100}{0.217}&\SEMIR{0.012}{0.157}&\SEMREL{0.072}{0.202}&\SEMIR{0.006}{0.135}&\SEMREL{0.082}{0.220}&\SEMREL{0.099}{0.215}\tableVerticalSeparator\\
&\multirow{2}{*}{Lyu~\cite{lyupan2013}}&\ASEMREL{0.010}{0.090}&\ASEMIR{0.002}{0.093}&\ASEMREL{0.002}{0.094}&\ASEMIR{-0.000}{0.089}&\ASEMREL{0.002}{0.091}&\ASEMREL{0.012}{0.097}\\
&&\SEMREL{0.007}{0.137}&\SEMIR{0.010}{0.157}&\SEMREL{0.009}{0.159}&\SEMIR{0.007}{0.148}&\SEMREL{0.013}{0.156}&\SEMREL{0.018}{0.150}\tableVerticalSeparator\\
&\multirow{2}{*}{Mahdian~\cite{mahdian2009}}&\ASEMREL{0.046}{0.146}&\ASEMIR{0.005}{0.082}&\ASEMREL{0.039}{0.128}&\ASEMIR{0.005}{0.086}&\ASEMREL{0.036}{0.132}&\ASEMREL{0.055}{0.158}\\
&&\SEMREL{0.055}{0.171}&\SEMIR{0.023}{0.159}&\SEMREL{0.057}{0.183}&\SEMIR{0.014}{0.146}&\SEMREL{0.052}{0.180}&\SEMREL{0.067}{0.191}\tableVerticalSeparator\\
\cmidrule{1-2}
\multirow{6}{*}{\rotatebox{90}{CFA-based}}&\multirow{2}{*}{Bammey~\cite{bammey}}&\ASEMIR{0.007}{0.084}&\ASEMREL{\BEST{0.682}}{0.329}&\ASEMREL{\BEST{0.501}}{0.427}&\ASEMIR{0.023}{0.095}&\ASEMIR{0.029}{0.091}&\ASEMREL{0.133}{0.288}\\
&&\SEMIR{0.021}{0.153}&\SEMREL{\BEST{0.665}}{0.349}&\SEMREL{\BEST{0.491}}{0.429}&\SEMIR{0.018}{0.107}&\SEMIR{0.020}{0.100}&\SEMREL{0.128}{0.290}\tableVerticalSeparator\\
&\multirow{2}{*}{Shin~\cite{shin2017color}}&\ASEMIR{0.007}{0.101}&\ASEMREL{0.104}{0.166}&\ASEMREL{0.085}{0.172}&\ASEMIR{-0.002}{0.042}&\ASEMIR{-0.001}{0.043}&\ASEMREL{0.015}{0.109}\\
&&\SEMIR{0.004}{0.123}&\SEMREL{0.099}{0.171}&\SEMREL{0.084}{0.179}&\SEMIR{-0.005}{0.058}&\SEMIR{-0.006}{0.059}&\SEMREL{0.012}{0.114}\tableVerticalSeparator\\
&\multirow{2}{*}{Choi~\cite{choi}}&\ASEMIR{0.026}{0.025}&\ASEMREL{0.603}{0.203}&\ASEMREL{0.420}{0.208}&\ASEMIR{0.001}{0.002}&\ASEMIR{-0.001}{0.003}&\ASEMREL{0.156}{0.114}\\
&&\SEMIR{0.030}{0.018}&\SEMREL{0.575}{0.191}&\SEMREL{0.385}{0.210}&\SEMIR{-0.001}{0.002}&\SEMIR{0.001}{0.001}&\SEMREL{0.139}{0.116}\tableVerticalSeparator\\
\cmidrule{1-2}
\cmidrule{1-2}
\multirow{12}{*}{\rotatebox{90}{JPEG-based}}&\multirow{2}{*}{Zero~\cite{ZERO}}&\ASEMIR{0.000}{0.000}&\ASEMIR{0.000}{0.000}&\ASEMIR{0.000}{0.000}&\ASEMREL{\BEST{0.796}}{0.349}&\ASEMREL{\BEST{0.732}}{0.413}&\ASEMREL{\BEST{0.638}}{0.451}\\
&&\SEMIR{0.000}{0.000}&\SEMIR{0.000}{0.000}&\SEMIR{0.000}{0.000}&\SEMREL{\BEST{0.756}}{0.387}&\SEMREL{\BEST{0.708}}{0.421}&\SEMREL{\BEST{0.624}}{0.453}\tableVerticalSeparator\\
&\multirow{2}{*}{CAGI~\cite{CAGI}}&\ASEMIR{0.004}{0.045}&\ASEMIR{0.000}{0.027}&\ASEMIR{0.002}{0.033}&\ASEMREL{0.038}{0.077}&\ASEMREL{0.044}{0.080}&\ASEMREL{0.031}{0.071}\\
&&\SEMIR{0.003}{0.052}&\SEMIR{-0.000}{0.042}&\SEMIR{0.001}{0.044}&\SEMREL{0.023}{0.077}&\SEMREL{0.028}{0.082}&\SEMREL{0.021}{0.073}\tableVerticalSeparator\\
&\multirow{2}{*}{FDF-A~\cite{ADQ3}}&\ASEMIR{0.031}{0.139}&\ASEMIR{-0.004}{0.087}&\ASEMIR{-0.003}{0.085}&\ASEMREL{0.226}{0.242}&\ASEMREL{0.228}{0.249}&\ASEMREL{0.203}{0.244}\\
&&\SEMIR{0.014}{0.169}&\SEMIR{-0.015}{0.139}&\SEMIR{-0.017}{0.139}&\SEMREL{0.216}{0.265}&\SEMREL{0.216}{0.273}&\SEMREL{0.187}{0.264}\tableVerticalSeparator\\
&\multirow{2}{*}{I-CDA~\cite{ADQ2}}&\ASEMIR{-0.000}{0.000}&\ASEMIR{-0.000}{0.000}&\ASEMIR{-0.000}{0.000}&\ASEMREL{0.416}{0.417}&\ASEMREL{0.422}{0.407}&\ASEMREL{0.381}{0.407}\\
&&\SEMIR{-0.000}{0.000}&\SEMIR{-0.000}{0.000}&\SEMIR{-0.000}{0.000}&\SEMREL{0.423}{0.408}&\SEMREL{0.414}{0.414}&\SEMREL{0.385}{0.408}\tableVerticalSeparator\\
&\multirow{2}{*}{CDA~\cite{ADQ1}}&\ASEMIR{-0.001}{0.034}&\ASEMIR{0.000}{0.055}&\ASEMIR{0.000}{0.052}&\ASEMREL{0.485}{0.339}&\ASEMREL{0.474}{0.344}&\ASEMREL{0.401}{0.360}\\
&&\SEMIR{-0.004}{0.068}&\SEMIR{-0.003}{0.098}&\SEMIR{-0.005}{0.097}&\SEMREL{0.449}{0.351}&\SEMREL{0.442}{0.350}&\SEMREL{0.378}{0.354}\tableVerticalSeparator\\
&\multirow{2}{*}{BAG~\cite{BLK}}&\ASEMIR{0.000}{0.015}&\ASEMIR{0.006}{0.078}&\ASEMIR{0.009}{0.079}&\ASEMREL{0.232}{0.461}&\ASEMREL{0.229}{0.458}&\ASEMREL{0.171}{0.430}\\
&&\SEMIR{0.002}{0.029}&\SEMIR{0.025}{0.164}&\SEMIR{0.026}{0.164}&\SEMREL{0.227}{0.459}&\SEMREL{0.223}{0.455}&\SEMREL{0.161}{0.430}\tableVerticalSeparator\\
\cmidrule{1-2}
\multirow{6}{*}{\rotatebox{90}{Generic tools}}&\multirow{2}{*}{Noiseprint~\cite{noiseprint}}&\ASEMREL{\BEST{0.127}}{0.200}&\ASEMREL{-0.001}{0.069}&\ASEMREL{0.066}{0.149}&\ASEMREL{0.013}{0.087}&\ASEMREL{0.178}{0.248}&\ASEMREL{0.153}{0.230}\\
&&\SEMREL{0.108}{0.232}&\SEMREL{0.002}{0.114}&\SEMREL{0.060}{0.179}&\SEMREL{0.016}{0.140}&\SEMREL{0.138}{0.279}&\SEMREL{0.128}{0.261}\tableVerticalSeparator\\
&\multirow{2}{*}{ManTraNet~\cite{mantranet}}&\ASEMREL{0.049}{0.091}&\ASEMREL{-0.000}{0.040}&\ASEMREL{0.074}{0.169}&\ASEMREL{0.004}{0.023}&\ASEMREL{0.095}{0.164}&\ASEMREL{0.112}{0.169}\\
&&\SEMREL{0.032}{0.099}&\SEMREL{-0.004}{0.065}&\SEMREL{0.053}{0.165}&\SEMREL{-0.000}{0.043}&\SEMREL{0.086}{0.171}&\SEMREL{0.107}{0.176}\tableVerticalSeparator\\
&Self-&\ASEMREL{0.082}{0.323}&\ASEMREL{0.028}{0.261}&\ASEMREL{0.036}{0.270}&\ASEMREL{0.011}{0.262}&\ASEMREL{0.078}{0.335}&\ASEMREL{0.138}{0.370}\\
&-Consistency~\cite{selfconsistency}&\SEMREL{\BEST{0.154}}{0.429}&\SEMREL{0.077}{0.393}&\SEMREL{0.082}{0.403}&\SEMREL{0.060}{0.386}&\SEMREL{0.151}{0.440}&\SEMREL{0.246}{0.425}\tableVerticalSeparator\\
\bottomrule
\end{tabular}
}
\caption{Results of different state-of-the-art forensics tools on our six datasets. The methods, on the left, are grouped by categories.
The metrics we use is the Matthews Correlation Coefficient (MCC), explained in more details in Section~\ref{sec:metrics}. As a baseline, any random classifier is expected to yield a score of 0. The mean of the MCC scores over each image of the dataset, as well as the standard deviation in parentheses, are shown for the \textcolor{c1}{exogenous mask} and \textcolor{c3}{endogenous mask} datasets. Grayed-out numbers represent results of methods on datasets that are irrelevant to said methods.
}
\label{fig:table}
\vspace{-10pt}
\end{table*}

On the hybrid dataset, the scores of the specific methods are lower than on the specific datasets, except for Splicebuster. For the JPEG-based methods, this is explained by the fact that one sixth of this dataset does not feature JPEG compression traces. For the CFA and Lyu and Mahdian noise-based methods, this is made worse by the fact that JPEG compression alters the previous noise and demosaicing artefacts, as shown in Figure~\ref{fig:complete_pipeline_canon_1250}. In particular, CFA-based methods are notoriously weak on JPEG images, since JPEG compression removes the high frequencies, in which mosaic artefacts lie. This can be seen on the visualization of the results in Fig.~\ref{fig:experiment_buste}, where the CFA-based method Bammey cannot make any prediction on the hybrid image, where the main part and the forged region were compressed with quality factors of respectively 93 and 75. On the other hand, Splicebuster obtains a higher score on the hybrid dataset since the analysis of noise residuals co-occurences enables this method to detect traces in multiple steps of the camera processing chain. 
%More generally, we can see that scores of methods in the hybrid dataset are closely linked to their scores in the JPEG datasets, which is expected since JPEG artefacts are more prominent than CFA artefacts and raw noise level changes.

As the generic methods we tested are based on (translation-invariant) convolutional neural networks, they can only detect shifts in periodicity at the exact place where the shift occurs, i.e. at the very border of a forgery. As a consequence, they unsurprisingly fail to detect shifts in both the JPEG and CFA grids, mostly showing performances that are barely better than random in both datasets. This, in turns, leads to a more spectacular result: Noiseprint, Splicebuster and ManTraNet all perform much better than random on the CFA Algorithm dataset. As seen on the CFA Grid dataset, this cannot be explained by their detecting a shift in grids when it occurs, and can thus only be explained by their ability to detect changes in the demosaicing algorithms used. This possibility had not been considered by state-of-the-art algorithms since the 2005 paper by Popescu and Farid~\cite{popescu_cfa}, when demosaicing algorithms were simpler and thus easier to detect. %This is even more surprising as none of those methods were initially trained to do so. \marina{Splicebuster is not trained..}

Regardless the dataset considered, the scores obtained by all of the methods have a high standard deviation with respect to their mean value. This suggests that, given a dataset, the scores in each individual image are not concentrated around the mean but rather spread on a large range of values. Hence, even for methods having low scores, some good detections are likely to happen.

%Concerning the scores obtained for the endogenous and exogenous masks, we observe that even if some methods show a better performance on the endogenous masks, this difference is negligible when considering the high standard deviation values the scores have.  We can conclude that none of the considered methods is taking advantage of the fact that endogenous masks are constructed using the image semantics. \qb{commented: selfconsistency actually does}

%Many methods, both specific such as Splicebuster, CAGI and I-CDA, and generic such as Noiseprint and ManTraNet, display a significant gap between their weighted IoU and their binary IoU scores with globally or locally best thresholds. This shows that thresholding remains a very difficult topic in image forensics, which is made even harder by the fact that there is rarely any one good threshold, as evidenced by the gap between the global and local best thresholds IoU scores. On this matter, we can see that the use of N-Cuts segmentation with Self-Consistency clusters the results and creates an almost-binary heatmap that can be used to provide a better automatic detection with a simple threshold.

\section{Discussion}\label{sec:dicussion}

The results are not significantly different for most methods between the datasets with endogenous and extragenous masks, showing that it is possible to draw a mask directly from the image and still obtain a challenging dataset. Nevertheless, evaluation on both datasets can reveal the ability of some methods to perform content-aware localization, not to make a detection --~as endogenous masks are still selected randomly from many possible masks of the image~--, but to help contour it. This is seen with Self-Consistency, which is a method that yields significantly better results on the endomasks, probably due to the fact that they use the image content to segment the detection.

The goal of this evaluation is not to compare and rank different methods with a single number, but to offer a rigorous insight to characterise the capabilities of each method.
Knowing the kind of inconsistencies to which each forensic tool is sensitive helps understand and explain its detections in uncontrolled cases.% Furthermore, the analysis of the results obtained when multiple traces are affected can help explain the failure to detect forgeries even if the specific trace they search for has been affected. \qb{as discussed, keep it for the journal article with a stronger study on this point (jpeg robustness)}

Methods that focus on detecting specific traces are often opposed to more generic methods. However, these studies show the complementary and possible synergies between the two paradigms. In particular, by training methods to generically detect inconsistencies, new possibilities arise. For instance, results on the CFA Algorithm datasets showed that, even without explicitly training them, neural networks were sometimes able to detect changes in the demosaicing algorithm, a fact that is usually considered almost impossible, especially locally, except with the most basic demosaicing algorithms~\cite{popescu_cfa}.

Our experiments also reveal a problematic issue with many of the tested methods. Even though they can yield decent scores, the standard deviations of theses scores over all images of the same dataset is often very high. Even though algorithms perform well on many forgeries, they also often yield false positives that require interpretation to be distinguished from true detections, such as Bammey and Noiseprint in some datasets of the example image seen in Fig.~\ref{fig:table}. Such phenomenon is further evidenced in the supplementary materials. This is a critical point for many methods, as it makes them usable only to a trained eye.
% \begin{figure}[t]
% \centering
% \begin{subfigure}{0.3\linewidth}
% \includegraphics[width=\linewidth]{figures/mantranet_false/hybrid_a.png}
% \caption{Input}
% \end{subfigure}
% \hfill
% \begin{subfigure}{0.3\linewidth}
% \includegraphics[width=\linewidth]{figures/mantranet_false/mask_a.png}
% \caption{Mask}
% \end{subfigure}
% \hfill
% \begin{subfigure}{0.3\linewidth}
% \includegraphics[width=\linewidth]{figures/mantranet_false/mantranet_h.png}
% \caption{Output}
% \end{subfigure}
% \caption{Although ManTraNet correctly detected the forgery on the left of the image, the detections at the right could also be mistaken for forgeries, even more as those false detections often follow semantic content. Image r01474187t, hybrid dataset with exomask.}
% \label{fig:mantranet_bad}
% \end{figure}

\section{Conclusion}
Image forensics datasets are usually grouped according to forgery types (eg. splicing, inpainting, or copy-moves), and do not separate the semantic content from the actual traces left by the forgery.
In this paper, we proposed to remove the semantic value of forgeries and to focus only on the traces. We designed a methodology to automatically create image forgeries that leave no semantic traces, by introducing controlled changes in the image processing pipeline. We built datasets by focusing on noise-level inconsistencies, mosaic and JPEG artefacts, and conducted an evaluation of some image forensics tools using this dataset.

Although we focused on three kind of changes in the forgeries, the same methodology could be applied to more traces, including PRNU inconsistencies, multiple compression traces, or image manipulations such as resampling.
More generally, we can address all forgeries where two different camera pipelines are involved. This includes copy-move, splicing and some methods of inpainting. Further work will incorporate other traces, such as those left by synthesis methods.

Our method can transform automatically large sets of images into forged images with fully controlled tampering cues and no bias that might cause overfitting. Besides evaluation of existing image forensics tools, %we believe
this methodology could also be used to train %a
forgery detection methods, although care would be needed so as not to overfit if using the same methodology for both training and evaluation. %\marina{if we say that this database could be used to train forgery detection methods we should argue how overfitting could be prevented. }\qb{Done}

%\tina{The whole methodology lets us evaluate methods in a non-semantic way, and the difference between the two datasets lets us evaluate if the methods use content info or not. If the results of the two datasets are significantly different, then one can conclude that the algorithm uses the image content, borders for example. The opposite is not true: if the results are similar, it does not mean that the algorithm doesn't look at the content.
%The fact that we did not see any difference between exo and endo does not let us conclude that the algorithms are not content aware. 2 possible explanations: the algos don't use content, or the segmentation of the image is not very well done regarding the semantic borders on which the algorithms were trained on.}

%\tina{Future work: how to evaluate the inpaintings?}
%\marina{The proposed methodology addresses forgeries where two camera pipelines are involved. However, there are other type of forgeries in which the tampering is done by synthesis. Further work is required to incorporate this type of forgeries.}

%\clearpage
\begingroup
  \renewcommand\thefootnote{}\footnote{%
\begin{wrapfigure}[3]{l}{1.6\baselineskip}\vspace*{-\baselineskip}\includegraphics[height=2\baselineskip]{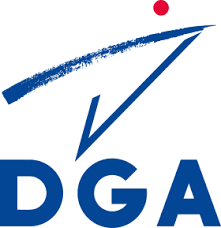}\end{wrapfigure}
Work partly funded by the French Minist\`ere des arm\'ees~--~Direction G\'en\'erale de l'Armement, and by grant ANR-16-DEFA-0004 Signature d'Images~--~ANR/DGA DEFALS challenge.
}%
  \addtocounter{footnote}{-1}%
  \endgroup
%%%%%%%%% BODY TEXT
{\small
\bibliographystyle{ieee_fullname}
\bibliography{biblio}
}

% \section{Snippets}

% Mention the fact that the dataset is asemantic but the segmentation (the mask creation) is done semantically.

% PRNU inchangé à mentionner.

% add figure with all the masks/heatmaps. Do it on one image hybridselect or on all the variants (from the datasets) of the image.

\end{document}